\documentclass[10pt,sigconf,letterpaper, nonacm]{acmart}

\usepackage[T1]{fontenc}
\usepackage[utf8]{inputenc}
\usepackage{microtype}
\usepackage[english]{babel}
\usepackage{blindtext}

\usepackage{amsmath,amssymb,amsthm,mathrsfs}

\usepackage{mathtools}

\usepackage{colortbl}
\usepackage{booktabs}
\usepackage{tabularray}
\usepackage{multirow}

\usepackage{graphicx}
\graphicspath{{figs/}}
\DeclareGraphicsExtensions{.pdf}
\usepackage{epsfig}
\usepackage{rotating}
\usepackage{wrapfig}
\usepackage{enumitem}
\usepackage{lipsum}
\usepackage{float}
\usepackage{booktabs}

\usepackage{tikz}
\usetikzlibrary{tikzmark}
\tikzset{mycircled/.style={circle,draw,inner sep=0.15em,line width=0.1em}}

\usepackage{algorithmic}

\usepackage[font=small]{caption}
\usepackage{subcaption}
\usepackage{textcomp}
\usepackage{xspace}
\usepackage{float}
\usepackage{url}
\usepackage[normalem]{ulem}

\usepackage{listings}
\usepackage{flushend}
\usepackage{pifont}
\usepackage{soul}
\usepackage{balance}
\usepackage{array}
\usepackage{capt-of}
\usepackage{gensymb}
\usepackage{footnote}
\usepackage[bottom]{footmisc}
\usepackage{siunitx}

\usepackage[normalem]{ulem}
\useunder{\uline}{\ul}{}
\hypersetup{draft} 
\usepackage{diagbox}
\usepackage{arydshln}
\usepackage{makecell}

\usepackage{xurl}

\usepackage{hyperref}
\hypersetup{final}

\makeatletter
\@ifpackageloaded{babel}
  {\newenvironment{nohyphens}
     {\par\sloppy\exhyphenpenalty=\@M
      \@ifundefined{l@nohyphenation}
        {\language=\@cclv}
        {\hyphenrules{nohyphenation}}%
     }
     {\par
      \@ifundefined{l@nohyphenation}
        {}
        {\endhyphenrules}%
     }
  }
  {
  }
\makeatother

\usepackage{xcolor}

\newcommand{\eg}{\textit{e.g.,}\xspace}

\newcommand{\ie}{\textit{i.e.,}\xspace}

\newcommand{\figs}{Figs.~}
\newcommand{\fig}{Fig.~}
\newcommand{\tab}{Table~}

\newcommand{\simpletitle}[1]{\noindent\textbf{#1}\xspace}

\DeclareTextFontCommand{\mytexttt}{\ttfamily\hyphenchar\font=45\relax}

\definecolor{Mycolor2}{HTML}{00FF00}

\definecolor{coralpink}{rgb}{0.97, 0.51, 0.47}
\definecolor{spirodiscoball}{rgb}{0.06, 0.75, 0.99}
\definecolor{turquoiseblue}{rgb}{0.0, 1.0, 0.94}
\definecolor{green(pigment)}{rgb}{0.0, 0.65, 0.31}
\definecolor{green(colorwheel)(x11green)}{rgb}{0.0, 1.0, 0.0}
\definecolor{limegreen}{rgb}{0.2, 0.8, 0.2}
\definecolor{fuchsiapink}{rgb}{1.0, 0.47, 1.0}
\definecolor{ceruleanblue}{rgb}{0.16, 0.32, 0.75}
\definecolor{lavenderindigo}{rgb}{0.58, 0.34, 0.92}
\definecolor{mangotango}{rgb}{1.0, 0.51, 0.26}

\renewcommand\footnotetextcopyrightpermission[1]{}
\setcopyright{none}

\acmDOI{}

\acmISBN{}

\acmPrice{}

\begin{document}
\title[5G in Mid-Bands: Europe and US]{Mid-Band 5G: A Measurement Study in Europe and US}

\author{Rostand A. K. Fezeu$^{\mathsection}$, Jason Carpenter$^{\mathsection}$, Claudio Fiandrino$^{\ddagger}$, Eman Ramadan$^{\mathsection}$, \\ Wei Ye$^{\mathsection}$,  Joerg Widmer$^{\ddagger}$,  Feng Qian$^{\mathsection}$, Zhi-Li Zhang$^{\mathsection}$}
\affiliation{%
 \vspace{-1em}\institution{University of Minnesota - Twin Cities, Minnesota, USA$^{\mathsection}$, \hspace{1em} IMDEA Networks Institute, Madrid, Spain$^*$}
}
\affiliation{%
  \institution{$^{\mathsection}$\{fezeu001,\hspace{0.2em}carpe415,\hspace{0.2em}ye000094, fengqian\hspace{0.2em}\}@umn.edu, \hspace{1em} $^{\mathsection}$\{eman,\hspace{0.2em}zhzhang\}@cs.umn.edu, \\ $^{\ddagger}$\{claudio.fiandrino,\hspace{0.2em}joerg.widmer\}@imdea.org}
}

\begin{abstract}
    Fifth Generation (5G) mobile networks mark a significant shift from previous generations of networks. By introducing a flexible design, 5G networks support highly diverse application requirements. Currently, the landscape of previous measurement studies does not shed light on 5G network configuration and the inherent implications to application performance. In this paper, we precisely fill this gap and report our in-depth multi-country measurement study on 5G deployed at mid-bands. This is the common playground for U.S. and European carriers. Our findings reveal key aspects on how carriers configure their network, including spectrum utilization, frame configuration, resource allocation and their implication on the application performance. 

\end{abstract}


\maketitle
\pagestyle{plain}

\section{Introduction}
\label{s:intro}

The commercial roll-out of Fifth Generation (5G) networks is nowadays widespread. Reportedly, the number of 5G subscriptions increased by 110 millions during the third quarter of 2022 and is projected to surpass the 5 billion barrier by the end of 2028~\cite{ericsson-report}. 3GPP has specified that 5G New Radio (NR) can operate at different radio bands for the radio access~\cite{ueRadio-part2}, \ie Frequency Range 1 (FR1) which includes low-bands (below $1$~GHz) and mid-bands ($1$ to $6$~GHz), and Frequency Range 2 (FR2) with high-bands at millimeter-wave frequencies (above $24$~GHz). 
Despite the much wider available bandwidth and early deployment around May 2019~\cite{Verizon-mmWave-5G}, mmWave 5G has largely been a disappointment, due to its short coverage range, sensitivity to blockage, wildly varying throughput and various other deployment challenges~\cite{narayanan2020firstlook,t-mobile-sidestepped-mmwave,mmWave-deployment-challenges,Forbes-reality-of-5G-in-US}. With little unused spectrum left in the low-bands, mid-bands currently offer the sweet spot for 5G deployments world-wide~\cite{mid-band-5G-sweet-spot}. 
According to the GSMA Intelligence report~\cite{gsma-report}, as of January 2023, only 7\% of the 5G worldwide roll-outs is in FR2, while the majority of the deployments is in mid-bands.

\simpletitle{Motivations and Main Goals of this Study.} Despite the fact that a tiny fraction of 5G roll-outs is in high-bands, the characterization of mmWave 5G performance 
has received comparatively more attention~\cite{narayanan2020firstlook,narayanan2020lumos5g,narayanan2021variegated,ramadan2021videostreaming,5g-meas-chicago-miami,narayanan2022mmavedepl,hassan2022vivisecting,fezeu20235gmmwave,liu2023unrealizedpotentials}
than 5G performance in mid-bands~\cite{xu2020understanding,5g-mmsys-europe,fiandrino2022uncovering,kousias2022nsa5g,parastar2023spotlight}. Further, while all the previous measurement studies in high-bands are performed in the US, those that focused on mid-bands  are scattered in various countries, \eg China~\cite{xu2020understanding}, Ireland~\cite{5g-mmsys-europe}, Spain~\cite{fiandrino2022uncovering}, Italy~\cite{kousias2022nsa5g}, and the U.K.~\cite{parastar2023spotlight}. In contrast to mmWave high-band 5G, 5G in mid-bands has been deployed in the US slower than in the rest of the world~\cite{Forbes-reality-of-5G-in-US}, partly due to the difficulty in acquiring the mid-band spectrum. While targeting 5G deployments in mid-bands from the onset, the landscape of 5G operators\footnote{Throughout the paper, we use carrier and operator interchangeably.} in Europe is, on the other hand, highly disaggregated vis-a-vis  the US counterpart. First, there are several national and cross-national carriers operating in the market and they usually own and run the network infrastructure in each European country \emph{independently}. 
Further, the European regulatory framework encourages fair competition among the national carriers and virtual operators and ensures agreements between carriers that enable consumers to seamlessly use the mobile device from any EU country at no additional cost irrespective of the home country. Based on such premises, \textit{our first main goal is to provide a comparative analysis of mid-band 5G configurations in the US and Europe} that sheds light on 5G mid-band deployments in each country.

Mid-band 5G is primarily deployed with Time Division Duplexing (TDD). The high demand for establishing services at such bands has led to a spectrum crunch and limited contiguous space for a single service. 
In contrast to Frequency Division Duplexing (FDD) which requires a part of downlink and uplink channels (and a "guard band" between them), 
TDD allows carriers to better utilize the spectrum resources with minimal fragmentation.
Furthermore, TDD makes it easier to deploy (massive) MIMO (multiple-input, multiple-output) which can significantly increase the downlink (DL) and uplink (UL) throughput performance with the same channel bandwidth. However,
proper configuration of the time allocation between UL and DL transmissions (\ie the frame structure) is key for efficient resource utilization and interference minimization with TDD synchronization.   Given the landscape of possible configurations, \textit{our second main goal is to investigate the implications of different deployment configurations on 5G network and application performance.}

\simpletitle{Challenges, Measurements and Specific Objectives.} Our objective in this paper is to contribute towards a holistic understanding of real-world 5G mid-band deployments, from network configurations all the way up to the application performance. Toward this goal, we conduct -- to our knowledge -- a first in-depth cross-continent, cross-country measurement study. The process of performing such measurements in the wild is non-trivial and faces several challenges. \textit{First}, the breadth of the study is huge: how to select countries, 5G operators to thoroughly survey several network configurations under limited manpower, resources and budget? \textit{Second}, how to orchestrate a cross-continental data collection to gather high-quality data? \textit{Third}, how to accurately profile network configurations and relate a wide range of (semi-)static and dynamic configuration parameters to the observed performance of 5G networks and applications running over them?

We tackle these challenges by setting up a comprehensive measurement platform comprising: \textit{(i)} multiple 5G smartphones -- thereafter referred to as UE (user equipment); \textit{(ii)} multiple SIM cards from a total of 9 different 5G operators in four European countries (\ie Spain, France, Germany, and Italy) and 3 major operators in the US,  \textit{(iii)} \textit{Accuver XCAL}~\cite{xcal}, a professional 5G measurement tool which collects detailed 5G NR Radio Access Network (RAN) protocol stack information;
\textit{(iv)} a set of diverse ``benchmarking'' applications such as iPerf, ping, file downloading with varying file sizes, and video streaming, and \textit{(v)} carefully selected servers that are located either within the operator's core networks or from three cloud service providers, Amazon, Google and Microsoft Azure. The measurement methodology and  measurement campaigns are presented in more detail in~\S\ref{s:measurement}. Our data collection spans more than 4500 minutes of network measurements ``in the wild", totaling 2+~TB data over 5G.

\begin{figure}[t]
    \centering
    \includegraphics[height=0.55\textwidth, width=0.47\textwidth, keepaspectratio]{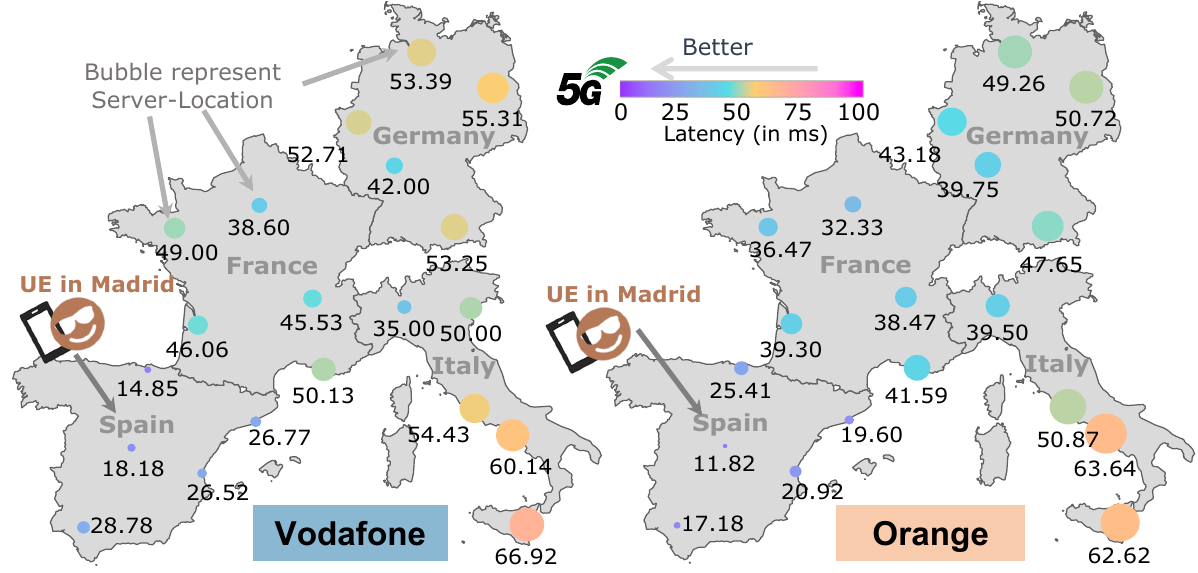}
    \vspace{-0.5em}
    \caption{\small Round-trip-time measurements from  UE in Madrid, Spain to Ookla speedtest servers located in four European countries. 5G carriers \& mid-band channels:  Vodafone Spain with 90 MHz (left) vs.  Orange Spain with 100 MHz (right). }
    \label{fig:orgSpO}
    \vspace{-1.8em}
\end{figure}

Leveraging our unique measurement dataset, we conduct a detailed analysis to obtain key insights regarding specific configuration settings such as channel bandwidth, frame structure (or frame TDD configuration), MIMO, carrier aggregation (CA), and many other parameters (see~\S\ref{s:config} and 
Tables~\ref{tab:euConfigs} \&~\ref{tab:usConfigs}), and their implications on  5G performance from a physical (PHY) layer perspective such as downlink and uplink throughput, user plane latency (see~\S\ref{s:phy-performance} and~\S\ref{s:furtherInvestigations}) as well as from the application perspective (see~\S\ref{s:app}). Fig.~\ref{fig:orgSpO} visualizes some sample round-trip-time (RTT) measurement results.

\simpletitle{Key Contributions and Findings.} We summarize our key contributions (marked with ``C'') and findings (``F'')  as follows:

$\bullet$ \textbf{C1.} We conduct the first large-scale, comprehensive measurement study of 5G in mid-bands in selected cities in four European countries as well as in the US. We uncover the 5G mid-band configurations used by the operators under study such as the channels/bands used, bandwidth, frame structures, MIMO layers, and so forth (see~\S\ref{s:config}). 

$\bullet$ \textbf{C2.} We identify the key factors that impact the observed mid-band 5G performance, including PHY downlink/uplink throughput and user-plane latency.  Not surprisingly, MIMO and carrier aggregation (CA) can significantly boost mid-band 5G PHY DL throughput: close to 1 Gbps (MIMO only) and 1.4 Gbps (MIMO+CA) in best cases (see~\S\ref{s:phy-performance}~\&~\S\ref{s:furtherInvestigations}).

$\bullet$ \textbf{F1.} While in theory, a wider channel bandwidth leads to higher (maximum) PHY throughput with more configured resource blocks (RBs), the observed PHY DL throughput does not correlate with channel bandwidth. For example, Vodafone Italy with a 80 MHz consistently outperforms Orange Spain with a 100 MHz channel under "good" channel conditions, while the latter performs better under "poorer" channel conditions. Among four operators with 90 MHz channels, performance varies depending on configuration parameters. 

$\bullet$ \textbf{F2.} As expected, across the board the PHY UL throughput is significantly lower than the DL throughput, and is on average far below 100 Mbps even under "good" channel conditions. With the non-stand-alone (NSA) deployment with dual connectivity, many operators also utilize
the 4G LTE connection for uplink transmissions -- especially when the 5G NR channel condition is  poor -- which complicates the 5G mid-band UL performance comparison. 

$\bullet$ \textbf{F3.} We find that the frame structures (TDD configurations) have a major impact on the observed mid-band 5G performance. This is particularly evident in the observed PHY user plane latency: The GSMA recommended 5G NR TDD configuration~\cite{gsma_2020} provides the lowest latency; it also yields additional benefits such as faster modulation and coding scheme (MCS) adaptation and thus lower block-level error rates (BLER). On the other hand, these benefits come with generally lower PHY DL throughput performance. 

$\bullet$ \textbf{F4.} The ``raw'' 5G PHY performance may not always translate to application QoE. We find that only when file sizes are sufficiently large (\eg $\geq$ 100~MBs), the faster PHY DL channels lead to significantly lower file download times. In terms of video streaming, while the 5G mid-band configuration matters, the stability of the channel environments and server placement also play an important role in determining the overall application QoE.

$\bullet$ \textbf{C3.} All in all, our study provides valuable insights into the roles of channel configurations in observed 5G mid-band performance. We believe our study will benefit not only the operators in better configuring and  optimizing their 5G mid-band services, but also users and applications in carrier selection, application tuning, and server placement choices.

$\bullet$ \textbf{C4.}
 Our work produces  the largest cross-continent, cross-country 5G mid-band measurement dataset that we know of. We plan to release the dataset and artifacts of our study publicly upon acceptance of this paper.
 
\simpletitle{Ethical Considerations.} The ethical considerations we have followed during this study can be found in Appendix~\ref{s:appendix-ethics}.

\section{Measurement Methodology}
\label{s:measurement}
In this section we present our measurement methodology along several dimensions: countries and 5G operators; measurement equipment and tools; benchmarking applications and measurement platform. Our measurement campaigns are meticulously orchestrated through the careful completion of three distinct tasks: First, comprehensive 5G scouting and deliberate carrier selection; second, measurement platform setup; and finally, the implementation of targeted measurement methodology for data collections.  We conclude with a brief summary of measurement campaigns and dataset.

\noindent
\textbf{Countries and 5G Operators.} Our study focuses on 5G mid-bands. In the US, multiple 5G bands are available: \textit{(i)}~high-bands (mmWave) (24.25–29.5 GHz, \eg band n261), \textit{(ii)}~mid-bands (3.3-3.8 GHz, \eg band n77), and \textit{(iii)}~low-band ($850$~MHz, n5)~\cite{narayanan2021variegated}. By contrast, in Europe carriers have deployed 5G only in mid (3.3-3.8 GHz, band n78) and low-bands ($700$~MHz, n28). To ensure an equitable evaluation, we conduct measurements in areas where only mid-bands are available. Hence, our study is conducted in selected cities spanning four EU countries (Spain, France, Germany, Italy), and the US. We aim to select operators with 5G mid-band services that have a major mobile market share of subscribers in the chosen countries~\cite{Taylor_2022, Davies_2017}. 
The carriers under study are Orange Spain and France (contract SIM cards), Vodafone Spain and Italy (contract SIM cards), Société française du radio téléphone (SFR) (prepaid SIM card),  Deutsche Telekom (contract SIM card) in Europe, and AT\&T,
T-Mobile and  Verizon in the US (multiple contract SIM cards for each US operator). For 5G scouting in each country, we use multiple resources such as relevant literature~\cite{fiandrino2022uncovering, narayanan2021variegated, narayanan2020lumos5g, ghoshal2022depth, 10.1145/3517745.3561465}, and platforms like Ookla speedtest~\cite{ookla_speedtest} and nperf~\cite{nperf}. In each chosen city and for each carrier, we pinpoint at least two locations with strong coverage and good propagation characteristics to carry out the measurements.

\simpletitle{Measurement Equipment and Tools.} 
To ensure a fair comparison, we use a consistent set of six flagship smartphones (Samsung Galaxy S21 Ultra, S21U) across all mobile operators in each country. 
We employ Accuver XCAL~\cite{xcal}, a professional tool, which provides  access to Qualcomm Diag (the diagnostic interface) to collect detailed 5G NR control plane and user/data plane information from the chipset. It runs on a laptop which is connected with up to six phones thereby enabling simultaneous data collection from all UEs at the different layers of the NR RAN protocol stack, from the PHY  layer to the Radio Resource Control (RRC) layer.

\begin{table}[t]
    \vspace{-1em}
        \captionsetup{font=small}
        \caption{Statistics of the data collected across countries.}
        \label{tab:dataStats}
        \centering
        \vspace{-1em}
        \setlength{\tabcolsep}{2pt}
        \resizebox{0.67\width}{!}{
        \centering
        \footnotesize
    \begin{tabular}{c c c c c c}
    \toprule 
    \multicolumn{6}{c}{\large \textbf{\LARGE Collected Dataset Settings and Statistics}} \\
    \midrule[.1em]
    \hline
    {\large Countries }
    & \large \textbf{Spain}
    & \large \textbf{France}
    & \large \textbf{Italy}
    & \large \textbf{Germany}
    & \large \textbf{USA}

    \\ \hline
    \multirow{3}{3cm}{\large Mobile Carriers} 
    & Orange & Orange & Vodafone & Telekom & T-Mobile\\

    & Vodafone & SFR & & Vodafone & Verizon\\

    &  &  &  &  & AT\&T
    \\ \hline
    \multirow{2}{*}{\large Server Platforms} & \multicolumn{5}{c}{\large Ookla Speed Test, On Premise}
    \\
    & \large Google
    & \large Azure
    &\large  Google
    &\large  AWS
    & \large AWS
    \\  \hline
    \large Nrof. Unique SIM cards & \multicolumn{5}{c}{\large 23} 
    \\
    \large Nrof. Smartphone (Models) & \multicolumn{5}{c}{\large 6 (3)} 
    \\
    \large Nrof. Mobile Operators & \multicolumn{5}{c}{\large 12} 
    \\

   \large Nrof. Servers Used & \multicolumn{5}{c}{\large 122} 
    \\
    \large Total Data consumed & \multicolumn{5}{c}{\large 2.02~TBs of data downloaded over 5G}
    \\
    \large Cummulative 5G Network Tests & \multicolumn{5}{c}{\large 4500+ minutes}
    \\
    \large Duration & \multicolumn{5}{c}{\large 15 Weeks}
    \\
    \bottomrule
    \end{tabular}
    }
    \vspace*{-0.5cm}
\end{table}

\noindent
\textbf{Benchmarking Applications and Measurement Platform.} 
In order to measure and evaluate mid-band 5G performance from both the \emph{network} and \emph{application} perspectives, 
we carefully select testing servers and run several ``benchmarking'' applications. For example, to test the ``raw'' network performance, especially the PHY downlink (DL) and uplink throughput and PHY user plane latency, we utilize Ookla Speedtest servers (and on-premise servers) that are located within the 5G operator network in the same city in which the test UE is located. We select only those 5G operators for which we have contract SIM cards to ensure our measurements will not be throttled. We conduct both download and upload iPerf experiments as well as Ping and traceroute testing to measure the DL/UL network throughput and latency. To evaluate the impact of 5G performance on application quality-of-experience (QoE), we design and conduct experiments using \textit{(i)} file downloads with different workload characteristics to mimic various mobile app behavior; and \textit{(ii)} video streaming.
We carefully select servers from 
the three dominant global cloud service providers, namely, Google Cloud Platform (GCP)~\cite{gcp},  Microsoft Azure Cloud~\cite{acp}, and Amazon AWS Cloud~\cite{AWS}, to run our experiments. We create a far-ranging measurement system that includes multiple back-end servers and a variety of customized tools and scripts used for both experiment automation and data collection, implemented on each individual server.
Clearly, conducting experiments \emph{in the wild} (as opposed to, a \emph{controlled} environment in a lab setting) poses significant challenges, as there are elements outside our control, such as channel propagation conditions, interference, and other users sharing the network resources. We aim to maintain consistency across multiple countries and ensure reliable data quality as much as possible, by carefully scouting the locations for experiments and conducting measurements under similar environments and settings whenever possible. When analyzing the collected data and comparing the performance of different operators across the countries/continents, we adopt the \emph{quasi-experimental design} (QED) approach~\cite{krishnan2013video} by controlling and conditioning on various parameters such as channel conditions, radio resource block (RB) allocation, and so forth (cf.~\S\ref{s:phy-performance} to~\S\ref{s:app}). While we have conducted mobility testing (where environmental factors are harder to control), the results presented in the paper are based solely on \emph{stationary} experiments to ensure fair and consistent evaluation and comparisons across 5G carriers and countries.

\noindent
\textbf{Measurement Campaigns and Data Collections. } 
Using the measurement tools and platform, we have carried out a rigorous \textit{cross-continent}, \textit{cross-country} measurement campaign. We travel and conduct experiments in different countries in a systematic manner.
After \textit{(i)} scouting 5G coverage in one metropolitan city per country, we \textit{(ii)} gather detailed logs via XCAL performing experiments at different days in the week and different times in a day.  We dedicate around 10 days for collecting measurements in each country. 
To summarize, we conducted more than 4500 minutes of measurements of 5G mid-band services offered by a number of operators in selected cities spanning four European countries as well as in the US over a period of several months (Oct. 2022 -- May 2023). The data usage across all carriers amounted to 2+~TBs. \tab\ref{tab:dataStats} provides an overview of the key settings of our measurement campaigns and statistics of dataset collected.

\begin{table*}[th]

\centering
    \begin{minipage}{0.6\textwidth}
      \centering
      \caption{[Europe] 5G mid-band Channel Configurations.}
      \label{tab:euConfigs}
      \vspace{-1em}
        \setlength{\tabcolsep}{3pt}
        \setlength{\extrarowheight}{3pt}
        \resizebox{0.48\width}{!}{
        \centering
    \begin{tabular}{|c||c|c|c||c|c||c||c|c}
    \hline
    \huge \textbf{\huge Country}
    & \multicolumn{3}{c||}{\huge \textbf{Spain}}
    & \multicolumn{2}{c||}{\huge \textbf{France}}
    & \multicolumn{1}{c||}{\huge \textbf{Italy}}
    & \multicolumn{2}{c||}{\huge \textbf{Germany}}
    \\ \hline \hline
    \large \textbf{Operators}
    & \multicolumn{2}{c|}{\textbf{\cellcolor{orange!12}\large Orange}} 
    & \textbf{\cellcolor{teal!30}\large Vodafone} 
    & \textbf{\cellcolor{orange!12}\large Orange} 
    & \textbf{\cellcolor{spirodiscoball!80}\large SFR} 
    & \textbf{\cellcolor{teal!30}\large Vodafone}
    & \textbf{\cellcolor{lime!50}\large Telekom} 
    & \textbf{\cellcolor{teal!30}\large Vodafone} 
     \\ \hline
    \textbf{SCS (kHz)} 
    & \multicolumn{8}{c}{\textbf{\cellcolor{gray!20}30}} 
    \\ \hline
    \textbf{Duplexing Mode} 
    & \multicolumn{8}{c}{\cellcolor{gray!20}TDD}
    \\ \hline

    \textbf{\shortstack{\footnotesize 5G NR Band}} 
    & \multicolumn{8}{c}{\cellcolor{gray!20}n78}
    \\ \hline
    \textbf{\makecell{\texttt{ChannelBandwidth} (MHz)}} 
    & \cellcolor{orange!12}100
    & \cellcolor{orange!12}90 
    & \cellcolor{teal!30}90 
    & \cellcolor{orange!12}90
    & \cellcolor{spirodiscoball!80}80 
    & \cellcolor{teal!30}80
    & \cellcolor{lime!50}90 
    & \cellcolor{teal!30}80 
    \\ \hline
    \textbf{\makecell{{Max. Bandwidth ($N_{RBs}$)}}} 
    & \cellcolor{orange!12}\textbf{273}
    & \cellcolor{orange!12}\textbf{245}
    & \cellcolor{teal!30}\textbf{245} 
    & \cellcolor{orange!12}\textbf{245}
    & \cellcolor{spirodiscoball!80}\textbf{217 }
    & \cellcolor{teal!30}\textbf{217}
    & \cellcolor{lime!50}\textbf{245}
    & \cellcolor{teal!30}\textbf{217} 
    \\ \hline
    \textbf{Carrier Aggregation} 
    & \multicolumn{8}{c}{\textbf{\cellcolor{gray!20}No}} 
    \\ \hline
    \textbf{\makecell{\footnotesize MIMO}} 
    & \cellcolor{orange!12}4
    & \cellcolor{orange!12}4 
    & \cellcolor{teal!30}4 
    & \cellcolor{orange!12}4
    & \cellcolor{spirodiscoball!80}4 
    & \cellcolor{teal!30}4
    & \cellcolor{lime!50}4 
    & \cellcolor{teal!30}4 
    \\ \hline
    \textbf{\shortstack{TDD Frame Structure}} 
    & \multicolumn{2}{c|}{\cellcolor{orange!12}DDDSU} 
    & \cellcolor{teal!30}DDDSU 
    & \cellcolor{orange!12}DDDSUUDDDD 
    & \cellcolor{spirodiscoball!80}DDDDDDDSUU 
    & \cellcolor{teal!30}DDDDDDDSUU 
    & \cellcolor{lime!50}DDDSU 
    & \cellcolor{teal!30}DDDSU  
    \\ \hline
    \textbf{\shortstack{Nrof. DL Sym. in `S'}}
    & \multicolumn{2}{c|}{\cellcolor{orange!12}10}  
    & \cellcolor{teal!30}10 
    & \cellcolor{orange!12}6 
    & \cellcolor{spirodiscoball!80}6 
    & \cellcolor{teal!30}6 
    & \cellcolor{lime!50}6 
    & \cellcolor{teal!30}10   
    \\ \hline
    \textbf{\shortstack{Nrof. UL Sym. in `S'}}
    & \multicolumn{2}{c|}{\cellcolor{orange!12}2}  
    & \cellcolor{teal!30}2 
    & \cellcolor{orange!12}4 
    & \cellcolor{spirodiscoball!80}4 
    & \cellcolor{teal!30}4  
    & \cellcolor{lime!50}2 
    & \cellcolor{teal!30}2 
    \\ \hline
      \end{tabular}
      }
    \end{minipage}\hspace*{-0.5cm}
    {\color{black}\;\vrule}
    \begin{minipage}{0.36\textwidth}
      \centering
      \caption{[US] Channel Configurations.}
      \label{tab:usConfigs}
      \vspace{-1em}
        \setlength{\tabcolsep}{6pt}
        \setlength{\extrarowheight}{3pt}
        \resizebox{0.48\width}{!}{
        \centering
    \begin{tabular}{c|c|c|c|}
    \hline
    \multicolumn{4}{||c|}{\huge \textbf{USA}}
    \\ \hline \hline
    \cline{1-3} 
    \multicolumn{2}{c|}{\textbf{\cellcolor{blue!20}\large T-Mobile} }
    & \textbf{\cellcolor{purple!20}\large Verizon}
    & \textbf{\cellcolor{yellow!40}\large AT\&T}
     \\ \hline
    15 & \multicolumn{3}{c|}{\textbf{\cellcolor{gray!20}30}} 
    \\ \hline
     \cellcolor{gray!20}FDD 
    & \cellcolor{gray!20}TDD 
    & \cellcolor{gray!20}TDD 
    & \cellcolor{gray!20}TDD 
    \\ \hline
    \cellcolor{gray!20}n25 
    & \cellcolor{gray!20}n41
    & \cellcolor{gray!20}n77 (C-band)
    & \cellcolor{gray!20}n77 (C-band)
    \\ \hline
    \cellcolor{blue!20}20+5
    & \cellcolor{blue!20}100+40 
    & \cellcolor{purple!20}60 
    & \cellcolor{yellow!40}40
    \\ \hline
    \cellcolor{blue!20} \textbf{51 + 11} 
    & \cellcolor{blue!20} \textbf{273 + 106}
    & \cellcolor{purple!20}\textbf{162}
    & \cellcolor{yellow!40}\textbf{106}
    \\ \hline
    \multicolumn{2}{c|}{\textbf{\cellcolor{gray!20}Mid- + Mid-Bands}} 
    & \textbf{\cellcolor{gray!20}Mid- + Low-Bands}
    & \textbf{\cellcolor{gray!20}Mid- + Mid-Bands}
    
    \\ \hline
    \cellcolor{blue!20}4
    & \cellcolor{blue!20}4 
    & \cellcolor{purple!20}4
    & \cellcolor{yellow!40}4
    \\ \hline
    \cellcolor{blue!20}N/A 
    & \cellcolor{blue!20}DDDSUUDDDD 
    & \cellcolor{purple!20}DDDSUUDDDD 
    & \cellcolor{yellow!40}DDDSUUDDDD 
    \\ \hline
    \multicolumn{2}{c|}{\cellcolor{blue!20}4}
    & \cellcolor{purple!20}6 
    & \cellcolor{yellow!40}6
    \\ \hline
    \multicolumn{2}{c|}{\cellcolor{blue!20}4}
    & \cellcolor{purple!20}4 
    & \cellcolor{yellow!40}4
    \\ \hline
      \end{tabular}
      }
    \end{minipage}
    \vspace{-1em}
  \end{table*}

\section{5G Mid-Band Configurations}
\label{s:config}

We provide an overview of {5G mid-bands} we studied across Europe and the US, including the
key configuration parameters used. 
We conclude by presenting the 3GPP (theoretical) maximum PHY throughput formula that ties together the roles of various 5G configuration parameters.

\subsection{Channel Bandwidth \& Frame
  Structures}   
Table~\ref{tab:euConfigs} summarizes the key 5G mid-band channel configuration parameters for several 5G operators in the four European countries under study (France, Germany, Italy, and Spain), while Table~\ref{tab:usConfigs} summarizes those of all three major 5G operators in the US. 
The details of how these parameters are extracted can be found in Appendix~\ref{ss:midBand-config}. Next, we discuss the implications of these parameters. 

\noindent
\textbf{5G Mid-Band Channel Frequencies and Bandwidth.}
From Table~\ref{tab:euConfigs}, we observe that the 5G mid-band channels used
by all operators in the four European countries fall within the n78 band (3.5 GHz, with the spectrum range of 3300 -- 3800 MHz), which is a sub-band of n77 (3300 – 4200 MHz), also referred to as the C-Band~\cite{cBand}. As specified by 3GPP~\cite{tddMode},
all n78 channels operate using TDD. This means that the \emph{downlink} (DL) data transmissions from the 5G base station (gNB) to the UE and the \emph{uplink} (UL) data transmissions from the UE to the 5G base station (gNB) are multiplexed in the same channel. The channel bandwidth used by each operator in each country varies from 80, 90, to 100~MHz (the maximum channel bandwidth for the n78 band).
All the n78 channels operate using the so-called 30~kHz subcarrier spacing (SCS), one of the 5G numerologies defined by 3GPP~\cite{3gppNumerology}. 
The channel bandwidth and SCS together determine the (maximum) \textit{Transmission Bandwidth Configuration} ($N_{RB}$) for the channel (see \fig\ref{fig:cbw_nrb} in Appendix~\ref{ss:midBand-config}).
$N_{RB}$ is specified in terms of \emph{resource blocks} (RBs) -- a resource block is the basic unit of radio resource allocation (in the frequency domain) by the gNB. 

Table~\ref{tab:usConfigs} shows that both the bands and channel bandwidth used by the three major US operators are more diverse in comparison to their European counterparts. Officially, both  AT\&T and Verizon use the C-band for their 5G mid-band deployments. As can be seen from the table, the channels used by both AT\&T and Verizon fall within the upper range of the n78 band, the same as those used in Europe: Verizon 5G mid-band channel bandwidth is 60 MHz, whereas AT\&T 5G mid-band channel bandwidth is only 40 MHz.\footnote{It is reported~\cite{signal-flash} that AT\&T has deployed another 3.45 GHz 5G mid-band channel of 40 MHz in Phoenix, US. However, this channel has not yet been deployed in the cities we have conducted our measurements.} In contrast, T-Mobile uses two different bands, n41 and n25 for its 5G mid-band deployments. In the n41 band~(2.5 GHz, 2496 -- 2690 MHz), T-Mobile utilizes two channels of bandwidth 100 MHz and 40 MHz, both operating in the TDD mode.
T-Mobile further utilizes two (pairs of) channels
of bandwidth 20 MHz and 5MHz each in the n25 band (1.9 GHz, 1850 -- 1915 MHz). The n25 band operates in the FDD (frequency division duplexing) mode, with a pair of DL and UL channels. T-Mobile utilizes these
n41 and n25 channels in various combinations for \emph{carrier aggregation} (see below).
Except for the n25 FDD channels, all other US mid-band channels use 30 kHz SCS; and the corresponding $N_{RB}$ is shown in  Table~\ref{tab:usConfigs}.

We remark that the 5G mid-band channels 
used by an operator are largely determined by the 5G mid-band spectrum they own or can acquire,
which is one of the biggest expenses in 5G deployments (see Appendix~\ref{ss:spain-use-study} for the Spain case study.)

\noindent
\textbf{Frame TDD Configurations.}
In both 4G LTE and 5G, each \emph{frame} is of length 10 ms, which is divided into 10 \emph{subframes} of 1~ms length. The subframes are further divided into \emph{slots} consisting of 14 symbols per slot. Unique to 5G NR, the number of slots per subframe hinges on the 5G numerology~\cite{3gppNumerology}. In the case of 30~kHz SCS, each subframe contains 2 slots of length 0.5ms~\cite{3gppNumerology}. As noted above, with the exception of T-Mobile n25 channels, all 5G mid-band channels in Europe and in the US use TDD. When data is transmitted in the DL and UL, it is determined by the so-called \emph{frame structure} or \emph{TDD configuration} (to be used interchangeably), which is also signaled as part of the 5G channel configuration parameters.

In contrast to 4G LTE which uses a set of 7 fixed  TDD configurations, 3GPP 5G specifications 
\emph{in theory} allow very flexible TDD configurations (and \emph{slot formats}) to support diverse emerging use cases, but \emph{in practice} only three TDD configurations are recommended for 5G mid-bands, due to spectrum harmonization, TDD synchronization and interference considerations~\cite{gsma_2020}. They are: \textbf{(1)}~\texttt{DDDSU}; \textbf{(2)}~\texttt{DDDDDDDSUU}; and \textbf{(3)}~\texttt{DDDSUUDDDD}, where ``D'' indicates a DL slot, ``U'' an UL slot, and ``S'' a special slot. The ``S'' slot  is specified by the slot format, often  containing some DL symbols and UL symbols separated by one or more ``guard period'' symbols (see Fig.~\ref{fig:gp_slot} in Appendix~\ref{ss:tddSync}). Hence the DL data can only be transmitted  in the DL slots and the DL symbols in the ``S'' slot, and the UL data can only be transmitted  in the UL slots and the UL symbols in the ``S'' slot. 
The first TDD configuration is the ``preferred'' configuration for 5G NR, whereas the other two are recommended when there are 4G LTE services using the same/neighboring bands/channels~\cite{gsma_2020}. 

From Tables~\ref{tab:euConfigs} and~\ref{tab:usConfigs}, we see that only four operators in Europe use \textbf{(1)}, while all others use either \textbf{(2)} or \textbf{(3)}.
Note that the length of \textbf{(2)}~\texttt{DDDDDDDSUU} and \textbf{(3)}~\texttt{DDDSUUDDDD} is 5ms, twice that of \textbf{(1)}~\texttt{DDDSU} (2.5ms). (The relation between the two TDD configurations with reference to a typical 4G LTE TDD configuration is depicted in Fig.~\ref{fig:tdd_sync} in Appendix~\ref{ss:tddSync}.) Clearly, with 7 vs. 6 DL slots every 5ms, \textbf{(2)}~\texttt{DDDDDDDSUU} and \textbf{(3)}~\texttt{DDDSUUDDDD}  will likely yield higher (PHY) DL throughput than \textbf{(1)}~\texttt{DDDSU}.
In~\S\ref{s:furtherInvestigations}, we see that this is indeed the case in general. Furthermore, we see that the TDD configurations also have a significant impact on PHY latency as well as  MCS adaptation and block-level error rates (BLER).

\noindent
\textbf{Maximum Modulation Order and CQI-MCS Mapping.} 
Besides (semi-)statically signalled 5G configurations, there are several \emph{dynamically signalled} parameters, typically on a \emph{per-slot} basis, that are crucial to 5G performance. They are part of the DCI (downlink control information) contained in the beginning of each slot, as part of the Physical Downlink Control Channel (PDCCH), and sent by the 5G base station (gNB).
These parameters include the specific RBs (and the number of consecutive symbols)
allocated for the UE DL transmissions, the modulation order (QPSK, 16QAM, 64QAM and 256QAM) and code rate (represented by a 5-bit MCS index) used, among others (see Fig.~\ref{fig:UE-gNB-exchange} and Appendix~\ref{ss:UE-gNB-exchange} for more details). The MCS index is typically chosen based on the channel quality indicator (CQI) periodically fed back by the UE, and may be further adapted based on the sounding reference signals (SRS) transmitted from the UE to the gNB in the UL slots/symbols (see~\S\ref{s:furtherInvestigations}). CQI (4-bit) has a range of [1,15], with 15 indicating the best channel condition. 

Several 5G mid-band operators use 256QAM as the maximum modulation order. The mapping from CQI to MCS is determined by the DCI format used in each slot. DCI format 1\_1 indicates to the UE the use of the 256QAM MCS table for CQI-MCS mapping, whereas  DCI format 1\_0 (used, \eg when the channel conditions worsen~\cite{book-5g-bullet}) indicates the use of the 64QAM MCS table.
The data transmitted in a slot is referred to as a \emph{transport block} (TB), the size (\# of bits) of a TB  is determined by MCS. Hence given the same number of RBs allocated to the UE, (i) with the same MCS table, a higher MCS index produces a larger TB size, thus higher PHY throughput (\emph{per-slot}); and (ii) with the same MCS index, using 256QAM MCS table  yields higher PHY throughput than using the 64QAM MCS table.  As we shall see, these dynamically signaled parameters play a crucial role in the observed 5G mid-band performance.

\noindent
\textbf{MIMO and Carrier Aggregation.}
Given a channel bandwidth which determines the maximum number of RBs that can be allocated to the UE, there are two ways to further increase the PHY (DL/UL) throughput: multiple-input multiple-output (MIMO) and carrier aggregation (CA)~\cite{ueRadio-part2}. MIMO works by utilizing multiple antennas on both the gNB and UE sides~\cite{MIMO}. As shown in Tables~\ref{tab:euConfigs} and~\ref{tab:usConfigs}, all operators in Europe and US employ (up to) $4 \times 4$ single-user MIMO (SU-MIMO) with four layers for both DL and UL transmissions. In other words, data may be spatially multiplexed and transmitted  using four ``streams" simultaneously, thus increasing the throughput 4-fold.
The number of MIMO layers used is signaled in the DCI mentioned above, which depends on the information periodically fed back (together with CQI)  by the UE (see Appendix~\ref{ss:midBand-config}), based on the channel condition.

On top of MIMO, CA can further boost  throughput by aggregating multiple channels (each is referred to as a \emph{component carrier}).
In our measurement data, we only observe T-Mobile in the US performs CA using mid-band channels.\footnote{
It is reported~\cite{signal-flash} that AT\&T also performs CA using its n77/C-band channel of 40 MHz as the primary cell and another 3.45 GHz 5G mid-band channel of 40 MHz as the secondary cell in places where both channels are configured, yielding a total of an aggregated channel of 80 MHz. However, we do not have measurement data to support it yet.
As an aside, we note that currently CA is performed  on the DL channels only, although both T-Mobile and AT\&T are actively exploring CA on the UL channels to increase the overall uplink throughput~\cite{T-Mobile_2023, Alleven_2023}.
} 
Since the maximum channel bandwidth defined for the n41 band is 100~MHz, T-Mobile divides the 
contiguous 140~MHz spectrum it owns
into two channels of 100~MHz and 40~MHz. 
T-Mobile combines both channels via CA
(using the 100 MHz as the primary cell) to achieve an aggregated channel bandwidth of 140~MHz. T-Mobile also employs CA using the two n25 FDD channels. In addition, T-Mobile can also combine all four channels for CA, yielding an aggregated channel bandwidth of 165~MHz. In contrast, Verizon performs CA by combining the 60~MHz n77 (C-band) TDD channel with a low-band (n5) FDD channel with a channel bandwidth of 10~MHz, yielding an aggregated channel bandwidth of 70~MHz. 

\noindent
\textbf{Non-Stand-Alone (NSA) vs. Stand-Alone (SA).} We note that
all European operators, and AT\&T and Verizon in the US deploy their 5G mid-band services using the NSA mode only at the time of our measurement studies. T-Mobile in the US has both NSA and SA mid-band 5G deployments. In order to compare 5G mid-band services across Europe and the US, in this paper we focus on NSA deployments only. 5G services deployed using the NSA mode imply that data will be transmitted over both 5G NR and LTE user (data) plane connections (so-called \emph{dual-connectivity}). We find that all operators (almost exclusively) rely on the 5G NR channel (which provides far higher bandwidth than LTE) for DL transmissions, whereas for UL transmissions, many operators combine both 5G NR and LTE connections. This complicates our comparative study of 5G mid-band UL performance.

\vspace{-0.5em}

\subsection{Maximum PHY Throughput} 
\label{ss:Max_PHY_Tput}
To illustrate the roles of various configuration (and dynamic) parameters in 5G performance, we  present the 3GPP 5G NR formula~\cite{5G-throughput-formula} 
for the  (theoretical) \emph{maximum} PHY throughput. 

\vspace*{-0.5cm}
\begin{equation}\label{eq:1}
\begin{aligned}
\scriptsize
Max\_PHY\_Throughput\ [in\ bits\ per\ second\ (bps)] =  \\ 
\sum^{J}_{j=1} \left\{ \nu^{(j)}_{layers}\cdot Q^{(j)}_{MCS} \cdot{\frac{12N^{BW(j),\mu}_{RB}}{T^{\mu}_s}} 
\cdot    f^{(j)} \cdot (1-OH^{(j)}) \right\} 
\end{aligned}
\end{equation}
where $J$ is the number of component carriers when CA is used; for each component carrier $j$, $\nu^{(j)}_{layers}$ is the number of MIMO layers used; $Q^{(j)}_{MCS}$  is the maximum modulation order
(\eg $6$ for 64QAM and 8 for 256QAM); $\mu$ is the {numerology}; $T^{\mu}_s =10^{-3}/{14\cdot 2^{\mu}}$ is the average {OFDM} symbol duration in a subframe for
numerology $\mu$;  $N_{RB}^{BW(j),\mu}$ is the maximum $RB$ allocation 
for a channel with bandwidth $BW(j)$ for numerology $\mu$; $f^{(j)}$ is the scaling factor, and may take values 1, 0.8, 0.75, and 0.4, depending on the MIMO layers and maximum modulation order used by  each component carrier. 
The formula applies to both downlink and uplink transmissions, where $OH^{(j)}$ is the overhead value depending on downlink/uplink
and the frequency range of the channel.

\begin{figure*}[t]
\begin{minipage}{0.7\textwidth}
\centering
\subcaptionbox{\label{fig:eu_cqi_more_12}Under Good Channel Conditions: CQI$\geq$12.}{\includegraphics[width=0.45\textwidth, height=1.4in]{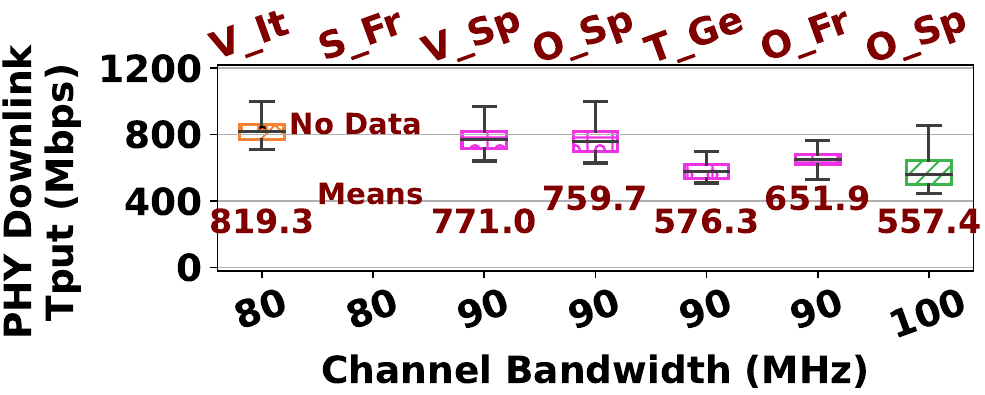}}
\hspace{0.05in}
\subcaptionbox{\label{fig:eu_cqi_less_10} Under Poor Channel Conditions: CQI$<$10.}{\includegraphics[width=0.45\textwidth, height=1.4in]{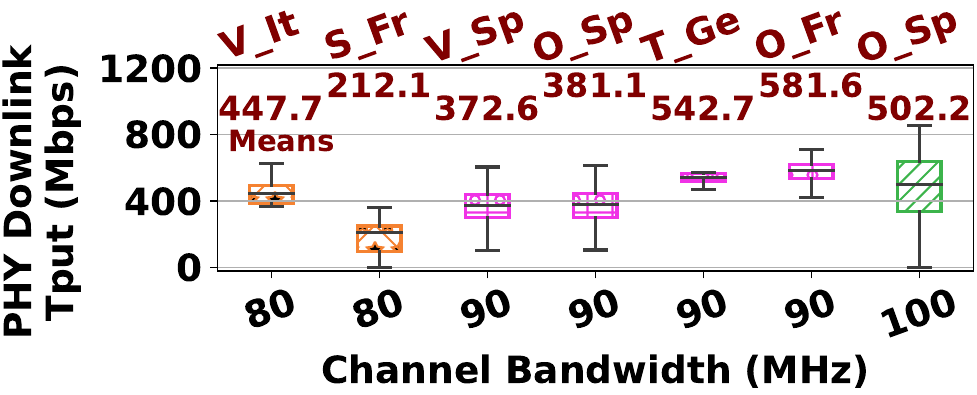}}
\vspace{-1em}
\caption{\small PHY DL Throughput of European Operators (with contract SIM cards): Vodafone Italy (V\_IT); SFR France (S\_FR) [no data under CQI $\geq 12$]; Vodafone Spain (V\_Sp); Orange Spain (O\_Sp); Deutsche Telekom (T\_Ge), Orange France (O\_Fr) and Orange Spain (O\_Sp)}
\label{fig:eu_dl_tput}
\end{minipage}
\hspace{0.05in}
\begin{minipage}{0.28\textwidth}
\centering
\includegraphics[width=\textwidth, height=1.4in]{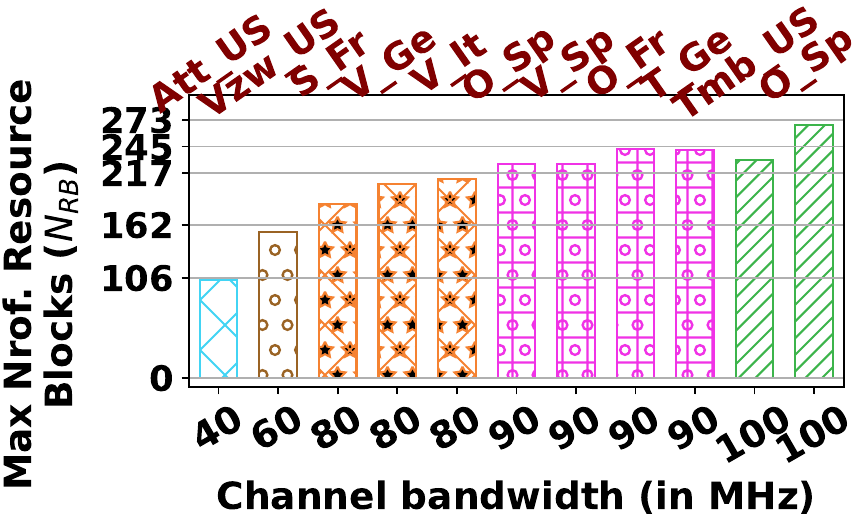}
\caption{Max Resource Block (RB) Allocation per operator and channel bandwidth.}
\label{fig:max_rb}
\end{minipage}
\end{figure*}

For all 5G mid-bands with SCS fixed to 30~kHz, $\mu=1$; for downlink, $OH^{(j)}=0.14$, and for uplink, $OH^{(j)}=0.08$.  When no CA is used (\ie $J=1$), $f^{(j)}=1$.  The number of RBs allocated per slot is bounded by $N_{RB}$ (Column 7 in Tables~\ref{tab:euConfigs}~and~\ref{tab:usConfigs}), namely,  $N_{RB}^{BW(j),\mu} \leq N_{RB}$. 
For a TDD channel, the TDD configuration  further constrains how many slots are available for DL vs. UL transmissions.  
For each frame (every 10~ms), there are 12~DL slots available for DL data transmissions using \texttt{DDDSU} vs. 14 DL slots available using 
\texttt{DDDDDDDSUU} or \texttt{DDDSUUDDDD}. In theory, the latter yields $\approx$1.17$\times$ higher PHY DL throughput than the former.\footnote{If we further consider symbols then 1.14$\times$, as \texttt{DDDSUUDDDD} has $7*14+6=104$ DL symbols, and \texttt{DDDSU} has $6*14+10=94$ DL symbols.} For UL data transmissions, both configurations produce 4 available UL slots for each frame (10~ms). As an example,
taking the frame TDD configurations into consideration, the equation (\ref{eq:1}) yields a  maximum PHY DL throughout of 1244~Mbps using \texttt{DDDSU} vs.\ 1374~Mbps using \texttt{DDDDDDDSU} for an 80~MHz channel.\footnote{Check the 5G NR throughput calculator tool~\cite{nr_tput_calculator} for more results.}
Besides the (semi-)static configuration parameters such as channel bandwidth ($N_{RB}$), maximum modulation order  and TDD configuration, the actual 5G PHY throughput performance will hinge critically on the various \emph{dynamic} parameters discussed earlier,  as will be shown in~\S\ref{s:phy-performance} and~\S\ref{s:furtherInvestigations}.

\section{5G Mid-Band PHY Performance}
\label{s:phy-performance}

\begin{figure*}[t]
\begin{minipage}{0.57\textwidth}
\centering
\subcaptionbox{\label{fig:us_noca}US Operators: when no CA is employed. }{\includegraphics[width=0.56\textwidth, height=1.3in]{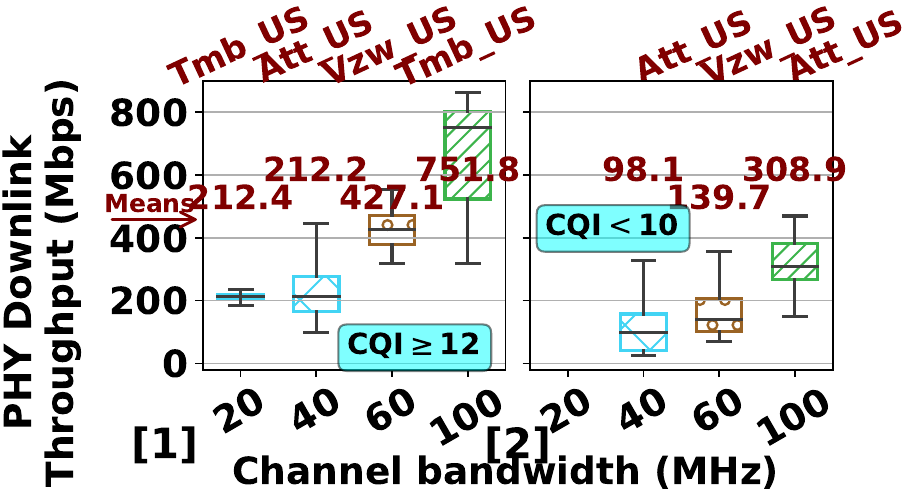}}
\hspace{0.04in}
\subcaptionbox{\label{fig:us_ca}Benefits of CA with TMB\_US.}{\includegraphics[width=0.32\textwidth, height=1.3in]{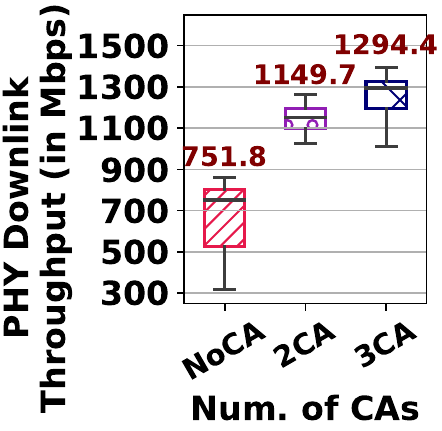}}
\vspace{-1em}
\caption{Effects of Carrier Aggregation.}
\end{minipage}
\hspace{0.03in}
\begin{minipage}{0.27\textwidth}
\centering
\includegraphics[width=\textwidth, height=1.3in]{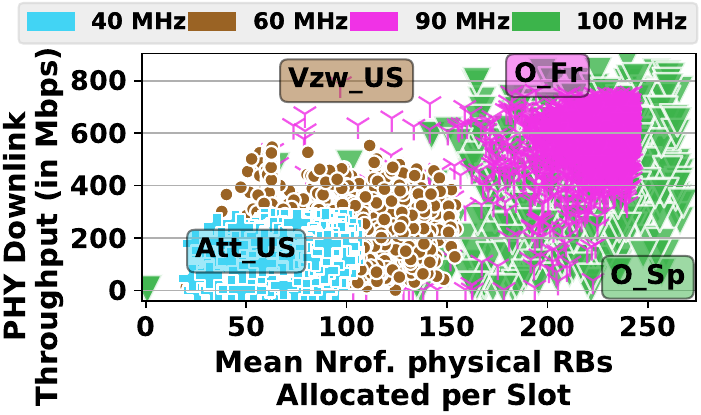}
\vspace{-1em}
\caption{Impact of Resource Block (RB) allocation on downlink throughput per operator.}
\label{fig:carrier_BW}
\end{minipage}
\hspace{0.05in}
\begin{minipage}{0.13\textwidth}
\centering
\includegraphics[width=0.7\textwidth, height=1.2in]{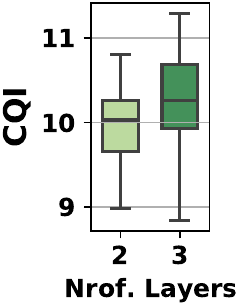}
\vspace{-1em}
\caption{Channel condition and MIMO layers.}
\label{fig:CQI_MIMO}
\end{minipage}
\end{figure*}

We now present a comparative study of the  overall 5G mid-band performance of
various operators in the four major European countries as well as in the US. 
In~\S\ref{s:furtherInvestigations}, we will further
investigate the joint effects of various configuration parameters on the observed 5G mid-band performance.

\subsection{PHY Downlink Throughput}
\label{ss:phyDL}

\fig\ref{fig:eu_cqi_more_12} shows the measured PHY DL throughput performance of the 5G mid-band channels of various European operators \emph{under generally ``good'' channel conditions, \ie $CQI \geq 12$.}
While in theory wider channel bandwidth should lead to a higher measured (maximum and average) PHY DL throughput as indicated by Eq.  (\ref{eq:1}), the observed performance is on the contrary. 
Vodafone Italy ({\tt V\_It}) with an 80 MHz channel outperforms 
all others, obtaining an average throughput of $\approx$ 819 Mbps and a maximum throughput close to 1 Gbps. Orange Spain (\texttt{O\_Sp}) with a 100 MHz channel has in fact the worst average throughput performance (557.4 Mbps). The performance of the four operators with a 90 MHz channel varies, with Vodafone Spain  (\texttt{V\_Sp}), for example, outperforming Deutsche Telekom (\texttt{T\_Ge}). 
When the channel conditions become "poorer" ($CQI < 10$), we see from  \fig\ref{fig:eu_cqi_less_10} that the performance of all operators suffers. While the average PHY DL throughput of  Orange Spain drops to 502.2 Mbps, that of Vodafone Italy with 447.7 Mbps is nearly halved.

\begin{figure*}[th]
\begin{minipage}{0.19\textwidth}
\centering
\includegraphics[width=\textwidth, height=1.3in]{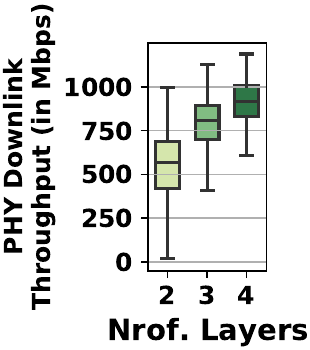}
\vspace{-1em}
\caption{Impact of MIMO on DL Tput.}
\label{fig:Tput_MIMO}
\end{minipage}
\begin{minipage}{0.54\textwidth}
\centering
\subcaptionbox{\label{fig:UL-Tput-good-CQI}Europe CQI$\geq$12.}{\includegraphics[width=0.49\textwidth, height=1.3in]{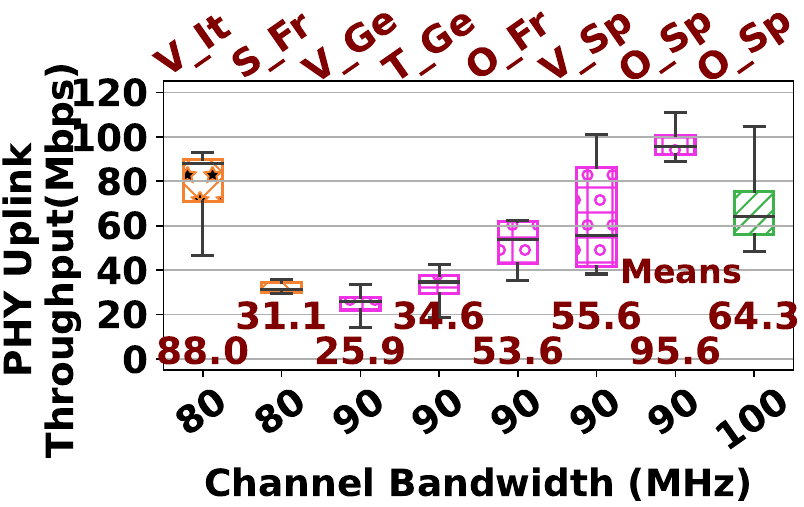}}
\subcaptionbox{\label{fig:UL-Tput-poor-CQI}Europe CQI$<$10.}{\includegraphics[width=0.49\textwidth, height=1.3in]{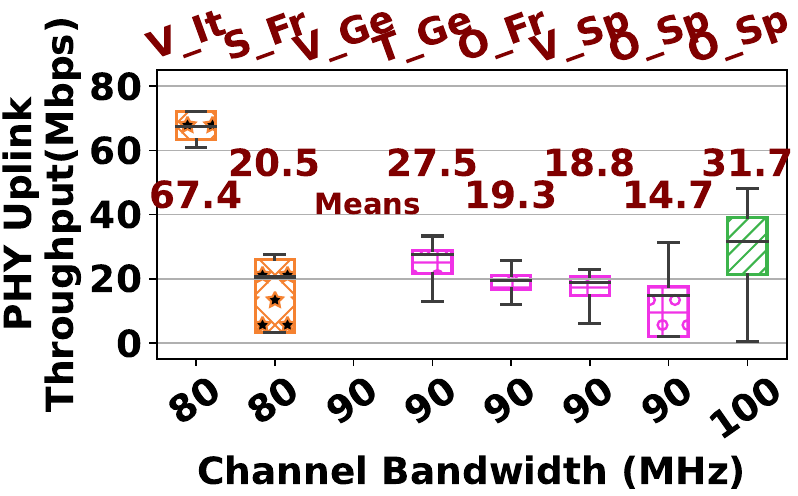}}
\vspace{-1em}
\caption{PHY Uplink Throughput.}
\end{minipage}
\hspace{.02in} 
\begin{minipage}{0.25\textwidth}
\centering
{\includegraphics[width=\textwidth, height=1.3in]{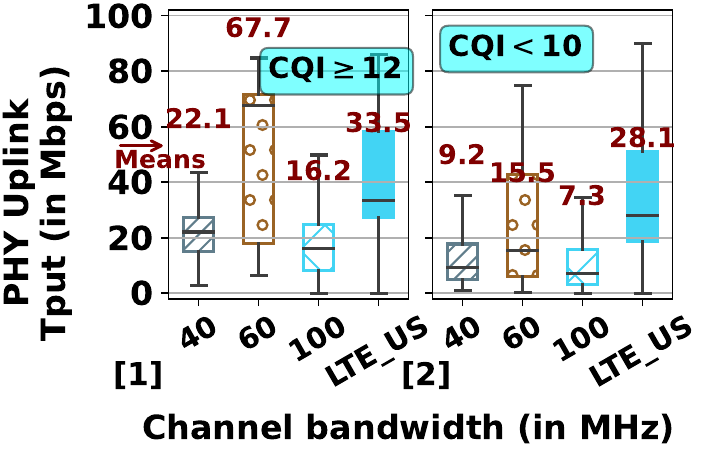}}
\vspace{-1em}
\caption{Phy Uplink Throughput US.}
\label{fig:ul_tput_us}
\end{minipage}
\vspace{-1em}
\end{figure*}

The performance of three major US operators under good (CQI $\geq$ 12) vs. poorer (CQI $<$ 10) channel conditions are shown in the left vs. right panel in~\fig\ref{fig:us_noca}, where we separately plot the performance of T-Mobile's 100 MHZ n41 channel and 20 MHz n25 channel \emph{with no carrier aggregation}. In the case of US operators, the performance in general conforms to the channel bandwidth, with  T-Mobile's 100 MHz channel performing the best and Verizon's 60 MHz channel performing second.
Under good channel conditions, the average PHY DL throughput of AT\&T's 40 MHz n77 TDD channel is roughly equal to that of T-Mobile's 20 MHz n25 FDD channel. Under poorer channel conditions, the performance of all three operators degrades also. (For T-Mobile's 20 MHz FDD channel, CQI always stays above 10 in the data we have collected.) Comparing the performance of the US operators and European operators, we see that T-Mobile's 100 MHz n41 channel outperforms that of Orange Spain's 100 MHz channel and is comparable to that of Vodafone Spain's 90 MHz n78 channel under good channel conditions. Under poorer channel conditions, T-Mobile's 100 MHz channel degrades similarly to that of Vodafone Spain's 90 MHz channel, but far worse than that of Orange Spain's 100 MHz channel. This can be partially attributed to the fact that poorer channel conditions lead to the reduced number of MIMO layers used, as shown in Fig.~\ref{fig:CQI_MIMO} and~\ref{fig:Tput_MIMO}, which demonstrates the effect of MIMO layers on the PHY DL throughput performance, using T-Mobile as an example.

T-Mobile is unique in its \emph{carrier aggregation}  (CA) capabilities. It can perform CA using (i) two of its n41 channels, yielding an aggregated channel bandwidth of 140 MHz, or (ii) both the n41 TDD channels and  n25 FDD channels, yielding, e.g., an aggregated channel bandwidth of 160 MHz with two n41 channels and one 20 MHz n25 channel. The PHY DL throughput performance of using these two CA options is shown in Fig.~\ref{fig:us_ca}. We see that CA can significantly boost the PHY DL throughput performance, with an average up to 1.3 Gbps and the maximum close to 1.4 Gbps.

To illustrate the relation between channel bandwidth and resource block allocation, in \fig\ref{fig:max_rb}, we plot the \emph{maximal} number of RBs allocated to the UE in a single slot by each operator. We see that nearly all operators may indeed allocate (close to) the maximum configured RBs in a channel to a single user (as given by $N_{RB}$ in Column 7 of Tables~\ref{tab:euConfigs} and \ref{tab:usConfigs}). \fig\ref{fig:carrier_BW} shows a scatter plot of some sample DL RB allocations by operators of differing channel bandwidth under various channel conditions. We see that in general wider channel bandwidth leads to higher resource block allocation.  We also find that the channel conditions do not affect the resource block allocation considerably. Hence resource block allocation alone cannot explain the varied performance we observe in \fig\ref{fig:eu_cqi_more_12}. In~\S~\ref{s:furtherInvestigations} we will dig deeper to explore the confluent effects of various configuration parameters.

\vspace{-1em}

\subsection{PHY Uplink Throughput}
\label{ss:phyUL}
We now shift our attention to the PHY UL throughput performance. 
Recall that currently no CA is used for UL data transmissions by any operator in the US or Europe.  \figs\ref{fig:UL-Tput-good-CQI} and~\ref{fig:UL-Tput-poor-CQI} show  the PHY UL throughput performance for the European operators under good channel conditions (\ie $CQI \geq 12$) vs. under poorer channel conditions (\ie $CQI < 10$).  Likewise, \figs\ref{fig:ul_tput_us}[1] and~\ref{fig:ul_tput_us}[2] show the PHY UL throughput results for the three US operators under good channel conditions (\ie $CQI \geq 12$) vs. under poorer channel conditions (\ie $CQI < 10$).
We see that in both Europe and the US, the PHY UL performance is significantly lower than that of the PHY DL performance. Even under good channel conditions, the maximum UL throughput of the \emph{best} operator (Orange Spain) barely reaches beyond 100~Mbps, and the average UL throughput of most operators falls between 20~Mbps to 80~Mbps. Under poorer channel conditions (CQI < 10), the average throughput of nearly all operators falls below 40 Mbps (with the exception of Vodafone Italy).

It is particularly surprising to note that T-Mobile's 100~MHz channel yields the worst performance. This seems to be puzzling. 
More in-depth analysis reveals that with the NSA mode, all three US operators prefer to utilize the LTE connection rather than the 5G NR connection for uplink transmissions, due to generally larger coverage and better channel quality. Using T-Mobile as an example, the blue boxplot at the rightmost side in each panel ([1] and [2]) of  \figs\ref{fig:ul_tput_us} shows the throughput of the LTE UL connection. We see that it is significantly higher than that of the 5G NR UL connection. We find that some of the European operators also resort to LTE for their UL transmissions, especially when the 5G NR channel conditions deteriorate. (Due to space limitations, we do not include any plot here.) As a result, the NSA dual connectivity makes the comparative analysis of the 5G mid-band PHY UL performance across the operators more challenging.

\begin{figure}[t]
    \centering
    \includegraphics[scale=0.53]{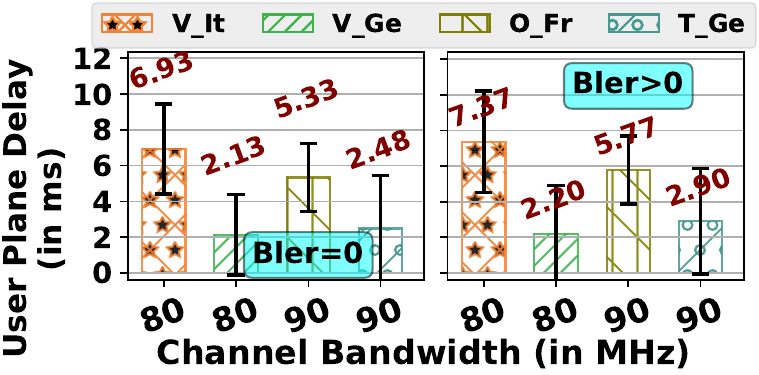}
    \caption{5G PHY User Plane Latency for different BLER.}
    \label{fig:PHY-latency}
    \vspace{-1.5em}
\end{figure}

\subsection{PHY Data/User Plane Latency}
\label{ss:phyLatency}
Finally, we compare the observed 5G mid-band PHY \emph{data (user plane) latency}  performance of the various European and US operators. Similar to~\cite{rossPam, Eiman_2020}, we define the user plane delay as -- \textit{the UL and DL latency on the PHY layer}. 
User plane latency is lowest when the so-called HARQ (Hybrid Automatic Retransmission reQuest) process is not invoked, namely, the block-level error rate (BLER) is zero. In case of poor channel conditions (and ``mismatch'' in the MCS that is used) or non-zero BLER, the data bits in a PHY layer need to be re-transmitted in subsequent DL/UL slots (hinging on the gNB RB allocation), thus yielding higher latency.

\fig\ref{fig:PHY-latency} shows the measured  PHY data plane latency results for four representative operators in Europe for two cases: \textit{(i)} BLER = 0 which represents the best case scenario; and \textit{(ii)} BLER > 0 where at least one HARD retransmission is performed, thus increasing the PHY user plane latency. We observe that the channel bandwidth has no bearing on the observed PHY user plane latency. For example, both with 80~MHz and BLER = 0, Vodafone Italy has the worst latency (6.93~ms), while Vodafone Germany has the best latency (2.13~ms).
With HARQ retransmissions (BLER >0), latency increases. We find that more retransmissions lead to higher latency in general. In~\S\ref{s:furtherInvestigations}, we will show that TDD configuration is a major contributor to the observed latency results.

\section{Further Analysis of the PHY}
\label{s:furtherInvestigations}
In TDD setting like that of 5G NR in mid bands, PHY layer downlink and uplink transmission opportunities dictate the traffic generated by both the network and the UE. In this section, we further investigate the joint effects of various configuration parameters on the observed 5G PHY performance.

\begin{figure*}[t]
\begin{minipage}{0.3\textwidth}
\centering
\includegraphics[width=\textwidth, height=1.3in]{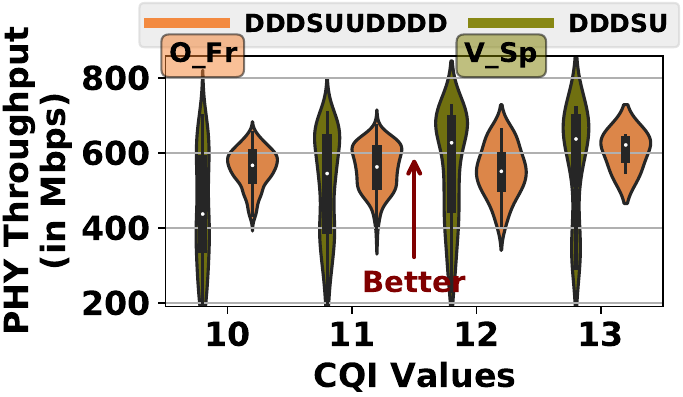}
\vspace{-1em}
\caption{[CB$=$90 MHz] DL Throughput vs CQI.}
\label{fig:cqi-tput}
\end{minipage}
\hspace{0.1in}
\begin{minipage}{0.3\textwidth}
\centering
\includegraphics[width=\textwidth, height=1.3in]{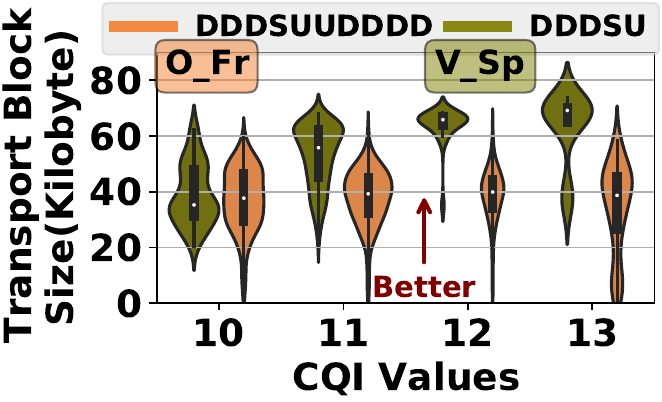}
\vspace{-2em}
\caption{[CB=90 MHz] Transport block sizes vs. CQI.}
\label{fig:cqi-tbs}
\end{minipage}
\hspace{0.1in}
\begin{minipage}{0.3\textwidth}
\centering
\includegraphics[width=\textwidth, height=1.3in]{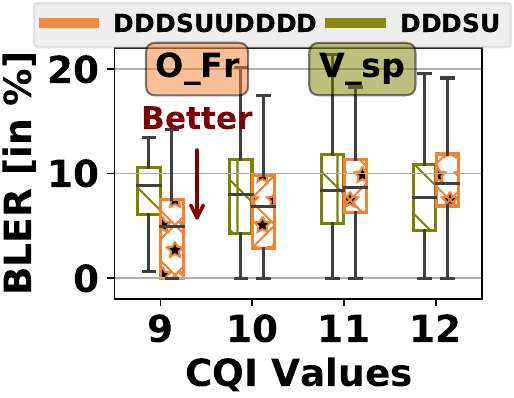}
\vspace{-2em}
\caption{[CB=90 MHz] BLER vs. CQI.}
\label{fig:cqi-bler}
\end{minipage}
\end{figure*}

\subsection{Impact on 5G PHY DL/UL Throughput}
\label{s:frameStructure-DL-UL-Tput}

\noindent\textbf{Frame Structure Impact.}
Recall Sec.~\ref{ss:Max_PHY_Tput}, the TDD configuration (frame structure) determines how many slots are available for DL and UL transmissions. For every frame (10~ms), the frame structure \texttt{DDDSUUDDDD} (or \texttt{DDDDDDDSUU}) can yield $\approx$1.17$\times$ higher (or 1.14$\times$ when considering symbols) PHY DL throughput than \texttt{DDDSU} in theory, assuming all other factors remain the same. We observe a ~1.6\% throughput gain in the DL with Orange France {\tt (O\_Fr)}, \texttt{DDDSUUDDDD} compared to Deutsch Telekom {\tt (T\_Ge)} with \texttt{DDDSU} (both have a 90 MHz n78 channel). However, the UL throughput is similar.

\noindent\textbf{MCS Impact.} Vodafone Spain attains a higher overall
PHY DL throughput performance than Orange France \emph{under favorable channel conditions}, but its PHY DL throughput performance is far worse \emph{under poor channel conditions} (recall Fig.~\ref{fig:eu_dl_tput}). This behavior can be attributed to a different factor. As mentioned earlier, the maximum modulation order used by Vodafone is QAM256, which employs a different MCS table from the one used by Orange France (and other operators), \ie QAM64. In other words, for QAM256, a different CQI-MCS mapping is used and the same modulation order (\ie the MCS index) typically yields higher transport block sizes, a higher target code rate, and better spectral efficiency. Fig.~\ref{fig:cqi-tbs} shows the transport block sizes for a selected set of CQI values for Orange France and Vodafone Spain. 

\noindent
\fig~\ref{fig:cqi-tput} shows the corresponding PHY DL throughput performance for the same set of CQI values. We can see that despite generally having fewer DL RBs allocated for DL transmissions (and similar UL RBs allocated for UL transmissions), the higher QAM/MCS order compensates this with higher transport block sizes, yielding overall better PHY throughput performance for Vodafone Spain over Orange France, \emph{when channel conditions are good} (e.g., with CQI $\geq11$).  
However, when channel conditions are poorer (e.g., with CQI $<10$), the higher QAM/MCS order becomes a liability due to potentially higher BLER. This is illustrated in Fig.~\ref{fig:cqi-bler}, where we compare the block-level error rates (BLER) of Vodafone Spain with the maximum modular order QAM256 vs. Orange France with the maximum modular order QAM64 under different CQI values. Consequently, poorer channel conditions lead to significantly worse PHY throughput performance for Vodafone Spain (see~\fig~\ref{fig:cqi-tput} with CQI $<10$).

\subsection{Impact on 5G PHY Latency}
\label{s:frameStructure-latency}
We ask why 3GPP recommends \texttt{DDDSU} as possible TDD configuration for 5G NR~\cite{gsma_2020} given that it provides fewer DL transmission slots (thus lower maximum DL PHY throughput). In fact, this is due potential latency improvements. The rationale is the following: with 2.5~ms \texttt{DDDSU} cycle period, the UE has the opportunity to sending the ACK/NACK HARQ feedback to the data transmitted in the D slots using the U slot within 2.5~ms. By contrast, the \texttt{DDDDDDDSUU} TDD configuration yields waiting times higher than 2.5 ms to provide HARD feedback for the data transmitted in the very first ``D'' slot. Likely, with the \texttt{DDDSUUDDDD} TDD configuration, the HARQ feedback for the data transmitted in the D slot right after the second U slot has to wait for 6 additional D slots before it can be sent; albeit the HARQ feedback for the data transmitted in the first 3 D slots can be sent within 2.5~ms.
This is indeed borne out by our measurement results.

\begin{figure}[h]
    \centering
    \includegraphics[width=0.47\textwidth]{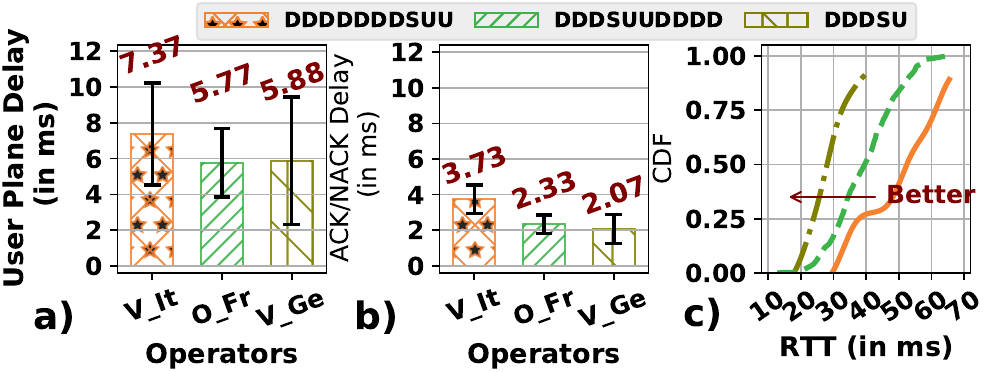}

    \caption{ Comparing a) user plane delay, b) ACK/NACK delay, and c) RTT with \texttt{DDDDDDDSUU}, \texttt{DDDSUUDDDD} and \texttt{DDDSU}. }
    \label{fig:phy_latency_kpi}
\end{figure}

Figs.~\ref{fig:phy_latency_kpi}(a) and (b) shows the overall PHY data/user plane latency and  the ACK/NACK delay (namely, the time lapse between the time the transport block containing user data  is received at the UE and the time the corresponding  ACK/NACK is sent), using the measurement data from Deutsch Telekom with a 90 MHz channel with \texttt{DDDSU} (V\_Ge in the plots), Orange France with a 90 MHz channel with \texttt{DDDSUDDDD} (O\_Fr in the plots), and Vodafone Italy with an 80 MHz channel with \texttt{DDDDDDDSUU} (V\_It in the plots).
We see that \texttt{DDDSU} indeed yields the best overall PHY latency due to the shortest ACK/NACK delay, \texttt{DDDDDDDSUU}  yields the worst overall PHY latency due to the longest ACK/NACK delay, whereas the PHY latency of \texttt{DDDSUUDDDD} falls in between. The PHY latency performance also translates to the end-to-end latency experienced by the applications, as evidenced by the round-trip-time (RTT) distributions shown in Fig.~\ref{fig:phy_latency_kpi}(c). 

\subsection{MCS Adaptation and BLER}
\label{ss:eff_and_rely}
\begin{figure}[h!]

\centering
\minipage{0.13\textwidth}
    \centering
    \includegraphics[width=\textwidth, height=1.2in]{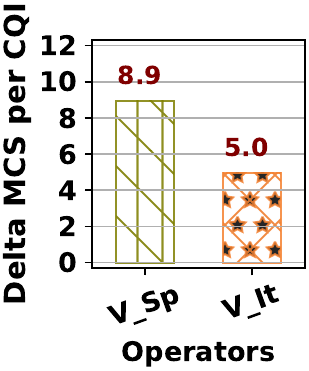}
    \vspace{-1em}
    \caption{MCS adaption.}
    \label{fig:mcs_adapt}
\endminipage
\hspace{0.1in}
\minipage{0.30\textwidth}
    \centering
    \includegraphics[width=\textwidth, height=1.2in]{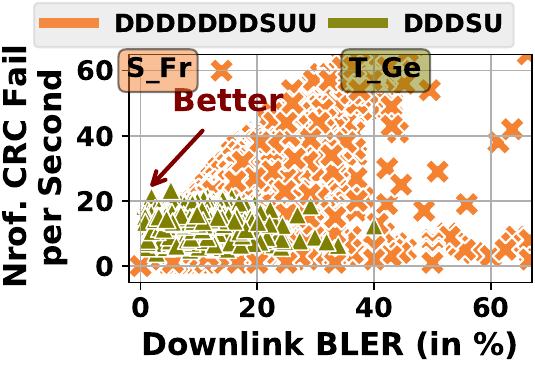}
    \vspace{-1em}
    \caption{Frame impact on BLER.}
    \label{fig:mcs_roc}
\endminipage
\end{figure}

The frame TDD configurations have another important effect on the MCS adaptation, and therefore the block-level error rate (BLER). These together will have an impact on the overall throughput (or rather, ``goodput'') and latency perceived by the upper layers and applications. Besides relying on CQI periodically reported by the UE to select the MCS index order for DL transmissions, the gNB also dynamically adapts it based on the sounding reference signals transmitted in the U slots. Hence, more frequent U slots lead to faster MCS adaptation. For example, \texttt{DDDSU} allows gNB to adapt the MCS every 2.5~ms in general, as opposed to every 5~ms when \texttt{DDDDDDDSUU} with maximum modulation QAM64. 
\fig\ref{fig:mcs_adapt} shows the average change rate of MCS value per CQI report period for an operator (Deutsch Telekom)  using \texttt{DDDSU} vs. an operator (SFR France) using  \texttt{DDDDDDDSUU}.

Adapting MCS more frequently enables the gNB to better cope with the varying channel conditions and, in turn, to improve BLER for DL transmissions. This behavior is made evident by fewer decoding errors at the UE that are quantified with more successful cyclic redundancy checks (CRC). \fig\ref{fig:mcs_roc} shows the relationship between the number of transport blocks that fail CRC ($\#TB_{fail}$) per second  and BLER ($\frac{\#TB_{fail}}{\#TB_{total}}$) per second for the same two operators as in Fig.~\ref{fig:mcs_adapt}. Clearly, \texttt{DDDSU} leads to fewer CRC fails and lower BLER.

\section{5G Application Performance}
\label{s:app}

Previous sections have revealed the impact of 5G carrier radio configurations on the PHY layer  performance. In this section, we further analyze the performance of 5G mid-band channels on applications' Quality of Experience (QoE). First, we conduct comparative profiling of varying application workloads on a target server to show the overall performance of different operators (\S\ref{ss:filedownload}). Second, we use video-on-demand (VoD) streaming to study the impact of channel bandwidth and frame structure on the application QoE for each operator (\S\ref{ss:video}). Finally, we study the impact of server location and cloud servers on the latency perceived by applications (\S\ref{ss:app_server}). Our main goal is to attempt to answer whether the 5G PHY performance impacts the applications' QoE or not.

\begin{figure*}[t]
\centering
\begin{minipage}{0.38\textwidth}
\includegraphics[width=\textwidth, height=1.3in]{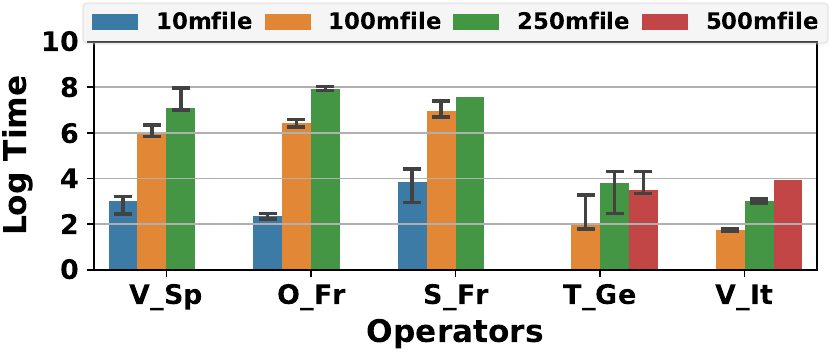}
\vspace{-2em}
\caption{File Download Time.}
\label{fig:fileDL}
\end{minipage}
\hspace{0.05in} 
\begin{minipage}{0.38\textwidth}
\centering
\includegraphics[width=\textwidth, height=1.3in]{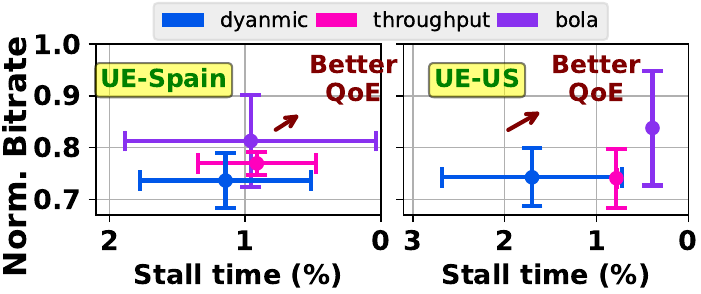}
\vspace{-2em}
\caption{BOLA Consistently Performs Better.}
\label{fig:bolaBetter}
\vspace{-1em}
\end{minipage}
\hspace{0.05in} 
\begin{minipage}{0.18\textwidth}
\centering
\includegraphics[width=\textwidth, height=1.3in]{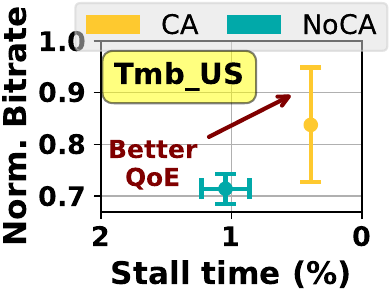}
\vspace{-2em}
\caption{Impact of CA on QoE.}
\label{fig:ca-vs-noca-qoe}
\end{minipage}
\hspace{0.1in} 
\end{figure*}

\begin{figure*}[t]
\centering
\begin{minipage}{0.35\textwidth}
\centering
\includegraphics[width=\textwidth, height=1.3in]{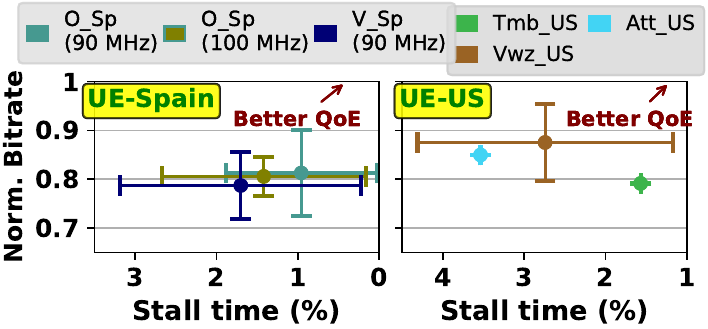}
\vspace{-1em}
\caption{Impact of Channel bandwidth on QoE.}
\label{fig:bitrate_stalltime}
\end{minipage}
\hspace{0.1in}
\begin{minipage}{0.2\textwidth}
\centering
\includegraphics[width=\textwidth, height=1.3in]{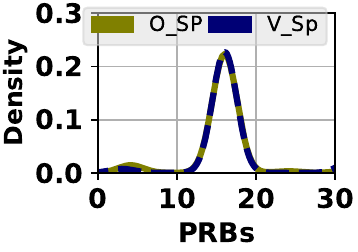}
\vspace{-1em}
\caption{PRBs utilization v.s. MCS in VoD.}
\label{fig:vod_prb_mcs}
\end{minipage}
\hspace{0.1in}
\begin{minipage}{0.35\textwidth}
\centering
\includegraphics[width=\textwidth, height=1.3in]{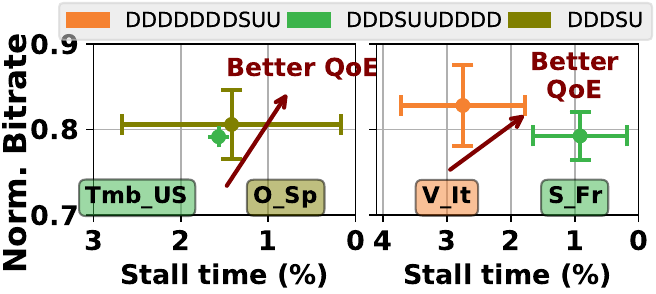}
\vspace{-1em}
\caption{Impact of Frame Structure on QoE, Left subgraph operators use 100~MHz, right subgraph use 80~MHz.}
\label{fig:qoeFrame}
\end{minipage}
\vspace{-1.5em}
\end{figure*}

\subsection{File Download Performance}
\label{ss:filedownload}
To mimic various mobile app behavior,  
we conduct a series of file download experiments using different file sizes in each country for a selected set of 5G operators. For the experimental results reported in this subsection, the files are stored on a target cloud server that is located in the same country where the UE is located. The server is selected in such a manner that: i)~it has sufficient processing capacity and the bandwidth along the Internet path is not a bottleneck; and 
ii)~the (measured) round-trip-times (RTTs) are consistently less than $\approx$50~ms 
Later in~\S\ref{ss:app_server}, we investigate the impact of server selection/placement on application performance. We perform a sequence of browser-based HTTP downloads at 2~s intervals to measure the file download times. We repeat each experiment multiple times. 

 \fig\ref{fig:fileDL} reports the file download time results for 5 operators in all four countries for file sizes of 10~MB, 100~MB, 250~MB, and 500~MB. The target server is in the same country as the UE. Note that the Y-axis is in the log 2 scale.  If we take the file size 250~MB, we see that Vodafone Italy and Deutsche Telekom have the best download time of less than 16~ms, while Vodafone Spain and the French carriers need around 128~ms -- 256~ms. Comparing these results to PHY DL throughput values from \fig\ref{fig:eu_cqi_more_12}, we find that Vodafone Italy has the highest throughput however this is not the case for Deutsche Telekom. Hence, we can conclude that the channel bandwidth and other 5G configurations are not correlated to the application QoE.

\vspace{-0.5em}

\subsection{Video Streaming Performance}
\label{ss:video}
For video streaming, videos are divided into several chunks of fixed length 2 -- 4 seconds encoded with different quality levels. We encode a 190-seconds custom video using \texttt{FFmpeg} with
\texttt{libx264} into seven qualities with different bitrates at 30 frames per second (fps) and 4~secs chunk length which is the suggested length for ABR~\cite{Lederer_2020}. The bandwidth required to download chunks from the lowest to the highest quality are $\approx$ 30~Mbps, 60~Mbps, 75~Mbps, 200~Mbps, 400~Mbps, 600~Mbps, and 750~Mbps respectively, selected based on the average throughput for operators which is $\approx$400~Mbps. Adaptive bitrate (ABR) algorithms are used to adapt the quality of the requested chunks according to the changing network conditions to improve QoE. For our experiments, we use an open-source video streaming system \texttt{DASH.js}~\cite{dash}. We test several popular ABR algorithms: (1)~BOLA~\cite{bola}, (2)~throughput-based~\cite{rate-based}, and (3)~dynamic bitrate algorithms implemented in \texttt{DASH.js}\footnote{We have also used L2A~\cite{l2a-mmsys-2020} and LoLP~\cite{lolp-tm-2022}, the results of which are not included in this paper.}.  We use a custom HTML-based player to measure the performance of each ABR algorithm across operators and cloud servers. Similar to~\S\ref{ss:filedownload}, for this section the target server used to host the video is in the same country as the UE. Our metrics include the normalized average bitrate\footnote{We normalize the bitrate to be between 0 and 1 using the bitrate of the highest available quality for the video (quality 7, which requires 750~Mbps).} of the downloaded video chunks  and stall time \% (\ie percentage of time spent while waiting for video chunks to be played). We note that, in our results, BOLA performs the best among all ABR algorithms we evaluated as can be seen in \fig\ref{fig:bolaBetter}, thus we resort to BOLA for our analysis.

\noindent
\textbf{Does channel bandwidth matter for QoE? }Recall that in~\S\ref{s:phy-performance}, we showed that the channel bandwidth determines the maximum number of resource blocks $N_{RBs}$ allocated to each UE. We also showed that $N_{RBs}$ is a major factor determining the PHY throughput and, in turn, the end-to-end throughput. Here, we examine whether it matters for application QoE, and in what conditions. For each country, we selected operators which have the same frame structure but different channel bandwidths. \fig\ref{fig:bitrate_stalltime} shows the normalized average bitrate (y-axis) and stall time \% (x-axis) when streaming using different operators in several countries.  We see that the difference in bitrates under different channel bandwidths is negligible for  orange Spain 90~MHz and 100~MHz channels. However, Orange Spain (90~MHz) performs slightly better than Vodafone Spain (90~MHz). For the US, Verizon (60~MHz) has the best bitrate while T-Mobile (100~MHz) has the smallest stall time. Hence, we can conclude that channel bandwidth does not have a clear impact on application QoE. 
For video streaming, downloading video chunks sporadically controlled by the ABR algorithm does not exhaust the network. We confirm this by showing in~\fig\ref{fig:vod_prb_mcs} the number of Physical Resource Blocks (PRBs) allocated by the network during streaming. Generally, we observe an average of 17  PRBs allocated for both operators {\tt (O\_Sp)} and {\tt (V\_Sp)}. Moreover, there is no need for the network to allocate the maximum number of RBs when streaming which is 245 for both operators. For the variance in bitrate and stall time, we attribute this to the changing channel conditions which causes BOLA to adapt when the throughput changes. 
Finally, \fig\ref{fig:ca-vs-noca-qoe} shows the performance for T-Mobile in US with and without CA. We can notice that CA helps increase the bitrate by around 14.72\% and minimize the stall time.

\noindent
\textbf{Does frame structure matter for QoE? }Another factor that affects PHY throughput is the frame structure as seen in~\S\ref{s:furtherInvestigations}. Here we examine its effect on Application QoE. \fig\ref{fig:qoeFrame} shows the normalized average bitrate (y-axis) and the corresponding stall time in \% (x-axis) comparing  different frame structures for a fixed channel bandwidth. We plot two scenarios: 1)~T-Mobile US (\texttt{DDDSUUDDDD}) and Orange Spain (\texttt{DDDSU}) with 100~MHz, and 2)~Vodafone Italy (\texttt{DDDDDDDSUU}) and SFR France (\texttt{DDDSUUDDDD}) both having 80~MHz (see Table~\ref{tab:euConfigs}). 
In general, some operators may have a better bitrate while others may experience less stall time. However, we can't say for sure that frame structure impacts the QoE as there are many factors that come into play that change the channel conditions, hence the throughput. This is further compounded by the adaptation of the ABR algorithms. Our insights for the previous sections were focused on PHY throughput while running iPerf experiments.

\subsection{Impact of Server Placement}\label{ss:app_server}
Lastly, we examine the implications of server placement on 5G application performance over 5G mid-band channels. For these experiments, we consider Europe only. Due to the relatively small size of each country compared to the US, major cloud service providers may not have cloud facilities in each country, or even have a presence in every country. Further, cloud capacities may vary from one country to another. Hence, server selection/placement is key to serve content to users effectively. For the results reported in this subsection, we conduct a series of experiments. Specifically, we use both file downloads and video streaming applications. For each country, the UE demands services placed on servers from two or more cloud service providers in different countries.

\begin{figure}[t]
    \centering
    \includegraphics[scale=0.45]{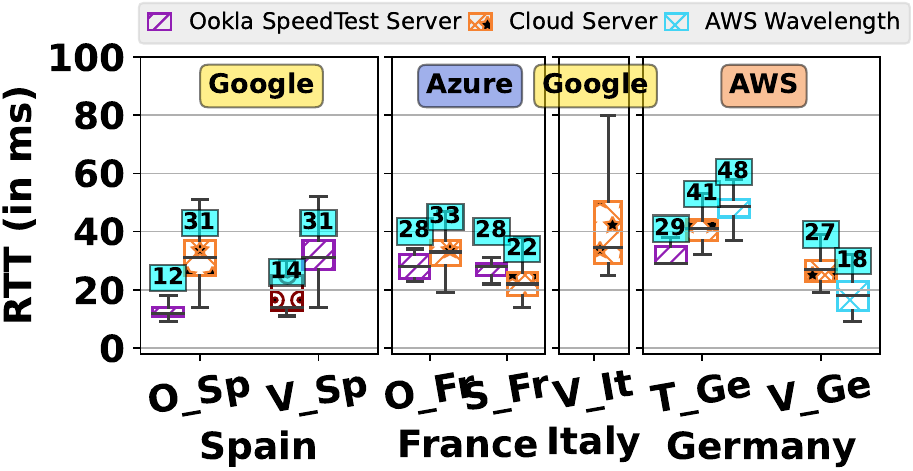}
    \caption{Latency Measurements to Different Servers.}
    \label{fig:rttEu}
    \vspace{-1em}
\end{figure}

Before reporting the server placement experimental results, we first show representative RTT measurement results (see Fig.~\ref{fig:orgSpO}) between the UE and target server. 
Using Google Cloud servers located in Spain and Italy, Azure Cloud servers located in France, and Amazon AWS Cloud (both Local Zone and Wavelength\footnote{
Amazon AWS Cloud features both so-called \emph{Local/Regional Zone} and \emph{Wavelength} servers in Germany. The latter service is in partnership with Vodafone Germany which provides direct connectivity to the computing facility via its 5G core network~\cite{AWS-GEM_Voda}. While we obtained permission to conduct RTT measurements on AWS Wavelength, we do not have the privilege to configure the servers for application performance experiments.}) servers located in Germany. For the RTT measurements, the UE and servers are located in the same country. The connection occurs over the different 5G operators available for that country. \fig\ref{fig:rttEu} shows representative results. For comparison, we also include the RTT measurements from the UE to Ookla speedtest servers that are \textit{often} located within the operator network in the same country. The claim is confirmed as the RTTs to the Ookla speedtest servers are \textit{generally} much shorter than to servers from the three cloud providers. 
(1)~In France, the RTT performance from the UE to the Azure Cloud server over SFR is better than that from the UE to the Ookla speedtest server. A careful examination of traceroute data reveals that in this case the Ookla speedtest server is not located within the SRF network. 
(2)~In Germany, the RTT performance from the UE to Amazon Wavelength server over Vodafone Germany is the best because the server has direct connectivity to Vodafone Germany's core network (see the previous footnote). By contrast, the RTT performance the UE observes to connect to the AWS Wavelength server over Deutsche Telekom is the worst. The results confirm that for server placement is key for RTTs. Owing to where cloud facilities are located and various Internet routing and other agreements among 5G operators and cloud providers, the RTT performance from the UE to a cloud server can vary significantly from 5G operator to 5G operator, cloud provider to cloud provider, and country to country.

\begin{figure}[t]
    \begin{minipage}{0.235\textwidth}
    \centering
    \includegraphics[width=\textwidth, height=1.3in]{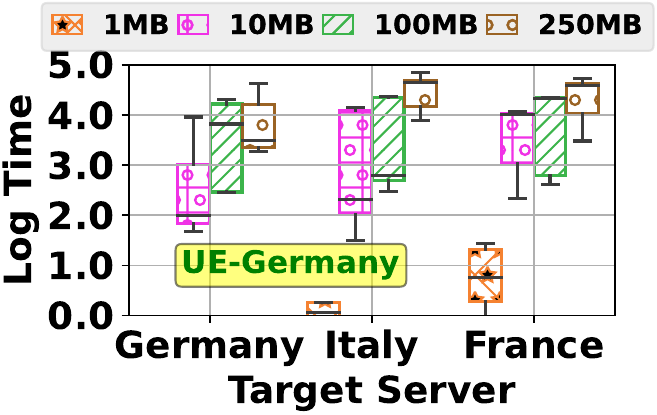}
    \caption{Impact of Server Placement on Download Time.}
    \label{fig:file-download-server-placements}
    \end{minipage}
    \hspace{0.01in}
    \begin{minipage}{0.2\textwidth}
    \centering
    \includegraphics[width=\textwidth, height=1.3in]{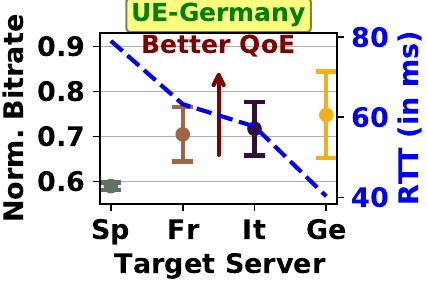}

    \caption{Impact of server Placement on VoD QoE.}
    \label{fig:video-streaming-server-placements}
     \end{minipage}
\vspace{-2em}
\end{figure}

We now show the impact of server placement on application performance over 5G. For this, we show the results with a UE located in Germany and using Deutsche Telekom to download a subset of files with different sizes (see \fig\ref{fig:file-download-server-placements}) and perform video streaming (see \fig\ref{fig:video-streaming-server-placements}) from several servers located in different countries. These are marked in \fig\ref{fig:video-streaming-server-placements} with RTT values according to their distance from the UE. Note that the Y-axis is in the log 2 scale in \fig\ref{fig:file-download-server-placements}. As expected, the download time increases as file size and distance between the UE and the target server increases. This is mirrored by the decreasing video streaming QoE as server distance increases.

Hence, the overall conclusion for this section is that there is no single factor that has a clear impact on application QoE compared to the PHY throughput we saw earlier. The perceived performance relies on the interplay between the logic of the application used, server location, and operator-cloud agreements, but not physical radio configurations. Moreover, currently, applications do not react to changes in the PHY layer, however, in the future cross-layer design may have an impact on application QoE. In addition, we currently do not have high-end applications which exhaust the available network bandwidth.

\section{Related Work}
\label{s:related}

There is a wealth of literature on measurement studies covering 5G since its commercial roll-out in 2019. The vast majority of the studies were conducted on 5G deployments (especially mmWave 5G) in the US and uncovered key aspects related to coverage, latency, throughput, and application performance~\cite{narayanan2020firstlook, xu2020understanding,5g-mmsys-europe,narayanan2020lumos5g,narayanan2021variegated,ramadan2021videostreaming,5g-meas-chicago-miami,narayanan2022mmavedepl,hassan2022vivisecting,fezeu20235gmmwave,liu2023unrealizedpotentials}. In contrast, the literature on measurement studies carried out in Europe is thin~\cite{5g-mmsys-europe,fiandrino2022uncovering,kousias2022nsa5g,parastar2023spotlight}. To the best of our knowledge, there exist no comprehensive and comparative studies on 5G in both the US and Europe, in particular with regard to the now widespread 5G mid-band deployments. To this extent, this paper fills an important gap. 

Some recent papers have investigated 5G network configuration parameters related to the management of high-band~\cite{narayanan2022mmavedepl} and mid-band~\cite{fiandrino2022uncovering} deployments. The breadth of this work surpasses that of~\cite{fiandrino2022uncovering} which relies on the open-source tool MobileInsight~\cite{mobileinsight}, as it encompasses a wider range of geographical regions and configuration settings.
The use of XCAL allows us to extract not only (semi-)static 5G configuration parameters, but also detailed (\emph{millisecond} scale) dynamic parameters exchanged between the 5G networks and UE.
Further, unlike~\cite{narayanan2022mmavedepl}, our work focuses on 5G mid-band. Other works aim at understanding throughput predictability at high-bands~\cite{narayanan2020lumos5g}, physical layer latency~\cite{fezeu20235gmmwave}, mobility management~\cite{hassan2022vivisecting}, power consumption~\cite{xu2020understanding,narayanan2021variegated} as well as  performance implications to specific applications like video streaming~\cite{narayanan2021variegated,ramadan2021videostreaming}, or to the broader application ecosystem from the user~\cite{liu2023unrealizedpotentials} and carrier perspective~\cite{parastar2023spotlight}.

\section{Conclusions}
\label{s:conclude}

We have presented, to the best of our knowledge, the  first large-scale cross-country, comparative 5G measurements study of 5G mid-band deployments in four major European countries and the US. Our study covers 9 European 5G operators and 3 major US 5G operators. Following a carefully designed measurement methodology, we have conducted extensive data collection campaigns totaling more than 2~TB data over 5G. Through \emph{in-depth} (at the \emph{millisecond slot} level),  \emph{cross-layer} data analysis,  we have uncovered the key 5G mid-band channels and configuration parameters employed by various operators in these countries, 
and identified the major factors that impact the observed 5G performance both from the network (physical layer) perspective  as well as the application perspective. Our results show the complexity and intricacies of 
 5G networks --- how various 5G configurations and dynamic parameters, such as channel bandwidth, resource blocks, channel conditions, modulation order, coding rate, MIMO, carrier aggregation, etc., conspire to influence the observed 5G network and application performance.
 Our datasets, analyses, and results (\eg roaming configurations and their impact on application performance) are richer and more expansive than what we have reported in this paper. Due to space limitations, we are not able to include them here. 

As 5G deployment in mid-bands has become dominant in the world, the results obtained in this work shed light on the roles of key 5G configuration parameters on network and application performance, and can aid 5G operators in better configuring and optimizing their 5G networks. Our study can also guide users, application developers and cloud providers in making carrier selection, application tuning and server placement choices. 
Our study also opens up new avenues for exploration.  For example,
our findings hint at the potential roles of  artificial intelligence and machine learning (AI/ML) in 5G networks: The intricate and inherently correlated factors, coupled with the highly variable performance of 5G, create opportunities for AI/ML in better configuring, optimizing and predicting 5G network and application performance. Last but not least, although our study focuses on mid-band 5G,  our methodology and analysis can be readily applied to low-band and high-band 5G networks. We believe that our study will provide invaluable insights into the design of next-generation (NextG) mobile networks and applications.

\bibliographystyle{ACM-Reference-Format}
\bibliography{reference}

\newpage
\appendix
\section{Ethical Considerations}
\label{s:appendix-ethics}
This study was carried out by the authors, paid, volunteer graduate and undergraduate students. We purchased several dedicated experiment-only smartphones and several unlimited plans from all mobile operators in this study. No personally identifiable information (PII) was collected or used, nor were any human subjects involved. Our study complies with the customer agreements of all 5G mobile operators and will not raise ethical issues.

\section{Extracting 5G Mid-Band Channel and Configuration Parameters}
\label{ss:midBand-config}

\noindent\textbf{Background about Initial Access Procedure.} 
The UE needs to obtain the master information block (MIB) and system information block (SIB) to connect to the network. 
The MIB contains the system frame number to start operating with the network (i.e., the \texttt{controlResrouceSetZero} and \texttt{searchSpaceZero}), and enables the UE to retrieve the allocated SIBs location (in RBs) by looking up Table 12$-$4 of 3GPP TS38.213 and Table 12$-$11 of 3GPP TS38.213, respectively~\cite{phyControl}. 

The SIB contains the cell's frequency and access-related information, permitting the UE to camp on. 

\begin{figure}[h]
    \centering%
    \vspace{-1em}
    \includegraphics[width=0.45\textwidth]{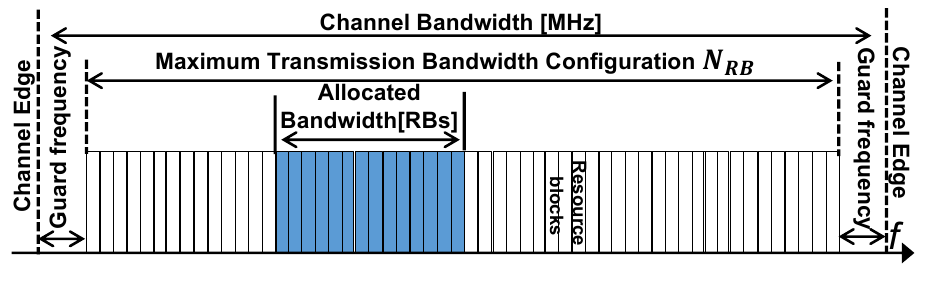}%
    \vspace{-1em}%
    \caption{Relationship between channel bandwidth and $N_{RB}$.}%
    \label{fig:cbw_nrb}%
\end{figure}

Among the cell information, \texttt{absoluteFrequencyPointA}, \texttt{offsetToCarrier}, and \texttt{carrierBandwidth} help UE identify the frequency channel resources: 
The \texttt{carrierBandwidth} retrieves channel bandwidth from the lookup table 5.3.2$-$1 \cite{ueRadio} in 3GPP TS38.101$-$1. 
The \texttt{absoluteFrequencyPointA} and \texttt{offsetToCarrier} determine the operating frequency channel. 
\fig\ref{fig:cbw_nrb} shows the relation between the channel bandwidth and $N_{RBs}$.

\begin{figure}[h]
    \centering%
        \includegraphics[width=0.37\textwidth]{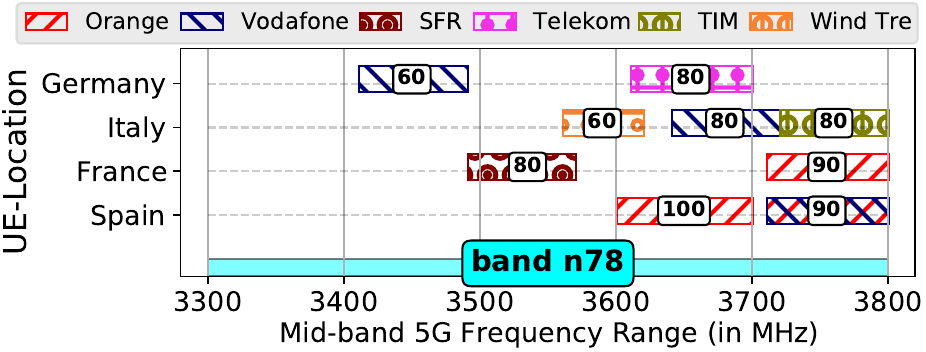}%
    \vspace{-1em}%
    \caption{[Europe] 5G Mid-band 5G channel bandwidth.}%
    \label{fig:spectrum-EU}%
    \vspace{-1em}
\end{figure}

\begin{figure}[h]
    \centering%
    \includegraphics[width=0.37\textwidth]{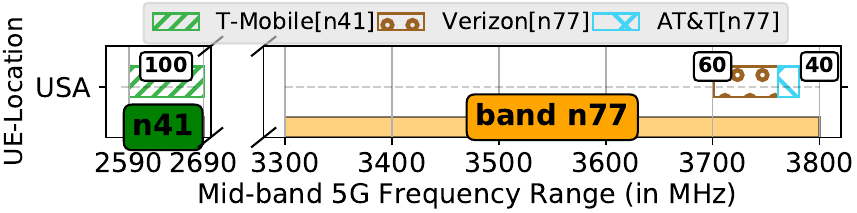}%
    \vspace{-1em}%
    \caption{[US] 5G channel bandwidth.}%
    \label{fig:spectrum-US}%
    \vspace{-1em}
\end{figure}

\noindent\textbf{Channel Information.} 
We retrieve those parameters via XCAL and locate the 5G mid-band channels used by each mobile operator. 
\fig\ref{fig:spectrum-EU} and \fig\ref{fig:spectrum-US} show the contiguous block of 5G mid-band channels observed in our dataset per country. This is a representation of all collected data. 
We find that Spain Orange and T-Mobile in the U.S. both own 100 MHz (the maximum channel bandwidth in 5G mid-band. Vodafone Spain, and Orange France own 90 MHz. SFR France, Telekom in Germany, and TIM, and Vodafone in Italy, all own 80 MHz. Vodafone Germany, Wind Tre in Italy, and Verizon in the U.S. use 60 Mhz, while AT\&T owns 40 MHz. Perhaps most interesting, we find that Orange and Vodafone in Spain share a 90 MHz channel bandwidth. As per the Mid-band 5G spectrum auction outcome in Spain~\cite{Mason_2020, Observatory_2019}, we conclude that Orange is using the Vodafone spectrum in our data. Nonetheless, we suspect that both MNOs are in a Reciprocal RAN Sharing agreement~\cite{Bourreau_2020}.

\section{CSI Feedback and DCI scheduling} 
\label{ss:UE-gNB-exchange}

\begin{figure}[h]
    \centering
    \includegraphics[width=0.45\textwidth]{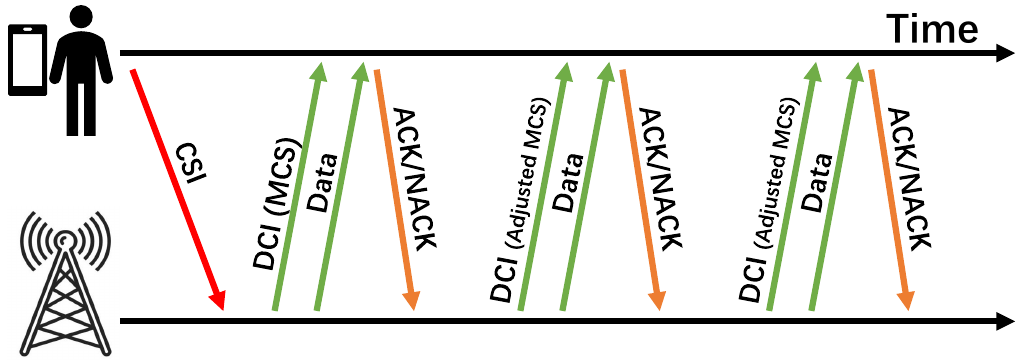} 
    \vspace{-1em} 
    \caption{CSI Feedback and DCI scheduling for downlink.} 
    \label{fig:UE-gNB-exchange} 
    \vspace{-1em}
\end{figure}

After access to the network, the UE will periodically (or upon request) transmit the current Channel State Information (CSI) to the gNB. This CSI feedback is averagely sent every tens of milliseconds. Based on the received CSI feedback, along with various other considerations such as traffic load, data queue status, UE priority profiles, and more, the gNB will dynamically determine the configuration for each data transmission (e.g., modulation order, coding rate, MIMO layers). The gNB will then convey this information to the UE using Downlink Control Information (DCI). The UE will provide feedback to the gNB regarding the status of the data packets by sending ACK/NACK feedback. The gNB will also refer to this feedback to adjust the data transmission decisions. Fig.~\ref{fig:UE-gNB-exchange} illustrates this communication procedure.

The DCI contains several key parameters, such as the number of layers, the number of RBs within the slot that have been allocated for the UE, and the  modulation and coding scheme (MCS) index used. 

The CSI feedback includes RI (rank indicator), PMI (precode matrix index), CQI (channel quality indicator), and LI (layer indicator), where the latter indicator depends on the previous. Specifically, RI indicates the number of independent data streams that can be simultaneously transmitted; PMI informs the optimal precoding matrix to be used; CQI reports the averaged channel quality and is the key indicator for gNB to decide the modulation and coding scheme (MCS); LI identifies the strongest layer. They altogether configure MIMO.

\section{Mid-Band Spectrum Acquisition: Spain Case Study} 
\label{ss:spain-use-study}

Besides the fact that 3GPP specifies the ``allowable'' channel bandwidths for each 5G band\footnote{For example, for the n78 band, the specified channel bandwidths are 10, 15, 20, 25, 30, 40, 50, 60, 70, 80, 90, and 100 MHz.}, the channel bandwidth used by operators is largely determined by the 5G mid-band spectrum they own or can acquire, \eg through spectrum auction, merger, or other means. Acquiring spectrum is one of the biggest expenses in 5G deployments. We conclude this section by using Spain as an interesting case study.

From the public information available on the Internet, the Spanish spectrum authority conducted two major auctions for the n78 band. The first auction, for the lower portion of the n78 band (3.4--3.6 GHz) for 5G services ended in 2016~\cite{Spain-n78-auction-1}: \textit{i)}~MasMovil (now part of Yogio Spain) acquired two 40 MHz channels/spectrum blocks, 3.4 -- 3.44 GHz and 3.5 -- 3.54 GHz, with a total of 80 MHz. \textit{ii)}~Telefonica acquired two 20 MHz channels, 3.44 -- 3.46 and 3.54 -- 3.56 GHz, with a total of 40 MHz. \textit{iii)}~Orange Spain acquired two 20 MHz channels, 3.46 -- 3.48 and 3.56 -- 3.58 GHz, with a total of 40 MHz. The second auction, for the upper portion of the n78 band (3.6--3.8 GHz) ended in July 2018~\cite{Spain-n78-auction-2}: \textit{iv)}~Vodafone Spain spent 198.1 million (m) euros (€) to acquire 18 5 MHz blocks for a total 90 MHz (3.71--3.8 GHz); and \textit{v)}~Orange Spain  spent €132.2m to acquire 12 5 MHz blocks for a total 60 MHz; and  vi)~Telefonica spent €107.4m to acquire 10 5 MHz blocks for a total 50 MHz. In addition, Spain also auctioned off 2 10 MHz channels\footnote{In the first auction in 2016, the Spanish spectrum authority reserved two 20 MHz channels in the 3.4 -- 3.6 GHz range: 3.48 -- 3.5 GHz; and 3.58 -- 3.6 GHz for the military radio location services. The two 10 MHz channels in the third auction are likely those in the 3.58 -- 3.6 GHz range, as they are now owned by Telefonica based on the information reported in~\cite{Spectrum-tracker-spain}.} in the n78 band in 2021~\cite{Spain-n78-auction-3}, with Telefonica and Orange Spain each bid for one of the two 10 MHz channels. The remaining 3.48 -- 3.5 GHz channel/spectrum block was also auctioned off and acquired by Yogio Spain.

As a conclusion, Orange Spain owns a total of  110 MHz in the n78 band, Telefonica 100 MHz, Vodafone 90 MHz, and Yogio 80 MHz. What is interesting is, based on the auction information, the n78 spectrum blocks acquired by Orange Spain and Telefonica are \emph{discontiguous},  scattered in the lower and upper portions of the n78 band. However, our measurement data indicates that Orange \emph{now} owns a \emph{contiguous} channel of 100 MHz (3.6 -- 3.71 GHz), which occupies the spectrum block that seems to have been acquired by Telefonica in 2018. We conjecture that Orange Spain and Telefonica must have formed an agreement to swap some of the n78 spectrum blocks they have acquired so that each owns a contiguous n78 5G mid-band channel. This is also confirmed by the information  in~\cite{Spectrum-tracker-spain}, which shows Telefonica owns a contiguous channel of 100 MHz (3.5 -- 3.6 GHz). 

Perhaps more interestingly, as indicated in Table~\ref{tab:euConfigs}, both Orange Spain and Vodafone use the same 90 MHz channel  (3.71 -- 3.8 GHz) for their respective mid-band 5G services. It appears that Orange Spain also ``owns'' the same channel that has been acquired by Vodafone Spain.
In fact, Vodafone is the sole owner of the spectrum block 3.71 -- 3.8 GHz (see~\cite{Spectrum-tracker-spain}). However, Orange Spain has an agreement with Vodafone Spain to share spectrum and the RAN infrastructure (and part of their core networks also)~\cite{Bourreau_2020}. This is the first instance of the so-called \emph{RAN sharing} (as specified  by 3GPP~\cite{ranSharing} which can take several forms)
that we have observed \emph{in the wild}.

\section{TDD Synchronization and 5G Frame Structures}
\label{ss:tddSync}
\begin{figure}[]
    \centering%
    \vspace{-1em}
    \includegraphics[width=0.45\textwidth]{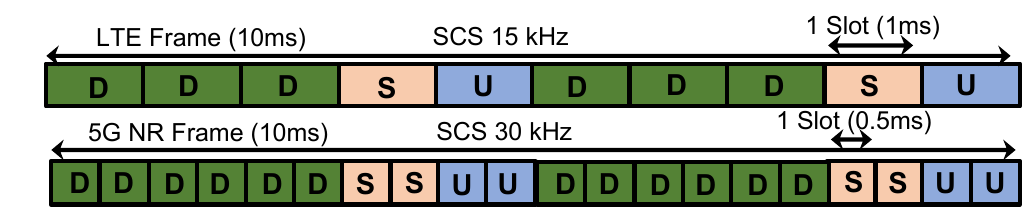}%
    \vspace{-1em}%
    \caption{Synchronization between TDD LTE DDDSU and 5G-NR DDDDDDDSUU Frames.}%
    \label{fig:tdd_sync}%

\end{figure}

Carriers configure specific frame structures generally to: 1) provide the
best compromise on performance in the presence or absence of incumbent systems like LTE, and  
2) to avoid interference in the presence of incumbent systems like LTE~\cite{gsmaTDD, ericssonTDD}. \fig\ref{fig:tdd_sync} shows an example of synchronization of the DL and UL transmission between \texttt{DDDDDDDSUU} and \texttt{DDDSU} frame structures when both LTE and 5G are deployed on the same/adjacent towers. Note that, the DL and UL transmission slots are aligned to minimize interference.  

\begin{figure}[]
    \centering%
    \vspace{-1em}
    \includegraphics[height=1.2cm, width=0.37\textwidth]{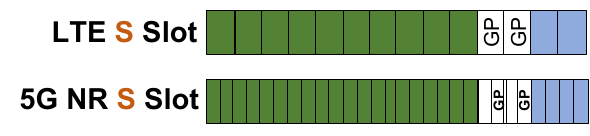}%
    \vspace{-1em}%
    \caption{Guard Periods (GP) configurations in the S slot in LTE and 5G NR frame.}%
    \label{fig:gp_slot}%

\end{figure}

Frame structures determines the DL/UL and UL/DL switching periods (SP). Guard periods (GP), configured in the 'S' slot are required at each switch for synchronicity and minimize interference, between transmission and reception. \fig\ref{fig:gp_slot} shows an example configuration of the 'S' slot with two GPs in LTE and four GPs in 5G NR. These GPs should be configured to allow sufficient time for \textit{i)} air propagation delays, \textit{ii)} time for the base station and UEs to change between transmission and reception modes, and \textit{iii)} enough time to correct cell phase synchronization errors. Note that, GPs is time when spectrum resources are not used. Thus, GP configurations should minimize this time while still allowing for the desired effect. 

\section{Additional Evaluations on RB Allocation}
\begin{figure}[h]
\begin{minipage}{0.24\textwidth}
    \centering
    \includegraphics[width=\textwidth, height=1.3in]{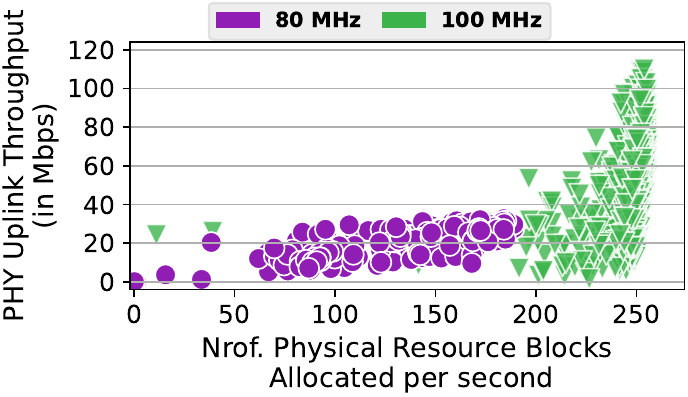}%
    \vspace{-1em}%
    \caption{PHY UL Throughput vs. $N\_{RBs}$.}%
    \label{fig:cbw_vs_PRBs_ul}%
\end{minipage}
\end{figure}

During our examinations into RB allocation, we also noted the UL throughput as it relates to RB allocation. Our results, as illustrated in \fig \ref{fig:cbw_vs_PRBs_ul} are fairly intuitive demonstrating a increase in RBs allocated generally on higher bandwidths.

\end{document}